\documentstyle[amssymb,pra,aps,epsfig,12pt]{revtex}
\newcommand{\be}{\begin{equation}}
\newcommand{\ee}{\end{equation}}
\newcommand{\br}{\begin{eqnarray}}
\newcommand{\er}{\end{eqnarray}}

\newcommand{\bd}{\begin{displaymath}}
\newcommand{\ed}{\end{displaymath}}

\newcommand{\bfig}{\begin{figure}}
\newcommand{\efig}{\end{figure}}

\def\3cdot{\cdot \cdot \cdot}

\def\om0{\omega _0}
\def\Om0{\Omega _0}

\def\text#1{{\rm{#1}}}

\def\->{\rightarrow}
\def\=>{\Rightarrow}
\def\-->{\longrightarrow}
\def\==>{\Longrightarrow}

\def\pr{^\prime}

\def\pr2{^{\prime\prime}}

\def\bfig{\begin{figure}}
\def\efig{\end{figure}}

\begin{document}
\title{{\Large Engineering Arbitrary Motional Ionic States Through Realistic
Intensity Fluctuating Laser Pulses }}
\author{R. M. Serra$^{1}$\thanks{%
Electronic address: serra@df.ufscar.br}, P. B. Ramos$^{1}$, N. G. de Almeida$%
^{1}$, W. D. Jos\'{e}$^{2}$, and M. H. Y. Moussa$^{1}$\thanks{%
Electronic address: miled@power.ufscar.br}.}
\address{$^{1}$ Departamento de F\'{\i}sica, Universidade Federal de S\~{a}o
Carlos,\\
PO Box 676, S\~{a}o Carlos, 13565-905, SP, Brazil. \\
$^{2}$ Departamento de Ci\^{e}ncias Exatas e Tecnol\'{o}gicas, \\
Universidade Estadual de Santa Cruz, Rodovia Ilh\'{e}us-Itabuna, km 16,\\
Ilh\'{e}us,\\
45650-000, BA, Brazil.}
\maketitle

\begin{abstract}
We present a reliable scheme for engineering arbitrary motional ionic states
through an adaptation of the projection synthesis technique for trapped ion
phenomena. Starting from a prepared coherent motional state, the Wigner
function of the desired state is thus sculpted from a Gaussian distribution.
The engineering process has also been developed to take into account the
errors arising from intensity fluctuations in the exciting laser pulses
required for manipulating the electronic and vibrational states of the
trapped ion. To this end, a recently developed phenomenological-operator
approach that allows for the influence of noise will be applied. This
approach furnishes a straightforward technique to estimate the fidelity of
the prepared state in the presence of errors, precluding the usual extensive 
{\it ab initio} calculations. The results obtained here by the
phenomenological approach, to account for the effects of noise in our
engineering scheme, can be directly applied to any other process involving
trapped-ion phenomena.

Journal-ref: Phys. Rev A {\bf 63}, 053813 (May, 2001) 
\end{abstract}


%
%%%%%%%%%%%%%%%%%%%%%%%%%%%%%%%%%%%%%%%%%%%%%%%%%%%%%%%%%%%%%%%%%%%%

\section{Introduction}

In recent years, experimental advances in the domain of cavity quantum
electrodynamics (QED) and trapped ions has motivated increasing interest in
the engineering of nonclassical states. The preparation of quantum states
constitute a crucial step towards testing fundamentals of quantum mechanics,
such as nonlocality and decoherence \cite{haroche}, and to the development
of devices such as logic gates for implementing quantum computation \cite%
{compq}. Theoretical schemes have been put forward for engineering arbitrary
states of the radiation field, both trapped in high-Q cavities \cite%
{vogel,k,eberly,g} and as a travelling wave \cite{pegg,dakna}, besides
electronic and vibrational states of trapped ions \cite%
{matos,kneer,buzek,solano,mc}.

In references \cite{vogel,k,eberly}, the desired superposition of
photon-number state in a cavity field is engineered from the vacuum state,
photon by photon. In Vogel {\it et al.}'s scheme \cite{vogel}, resonant
atom-field interactions are required to create a general single-mode
resonator, while Parkins {\it et al.} \cite{k} suggest a method using
adiabatic transfer of Zeeman coherence, and Law and Eberly \cite{eberly}
have provided an approach without the need to prepare multilevel electronic
superpositions. In reference \cite{g} the authors, despite starting from a
coherent state of the cavity field, do not provide a scheme for generating
an arbitrary superposition state. In the travelling wave domain, Pegg {\it %
et al.} \cite{pegg} proposed a method to generate an arbitrary superposition
of a zero- and one-photon field state, based on state truncation of
travelling optical fields. A similar scheme for generating arbitrary quantum
states of a travelling wave was proposed by Dakna {\it et al.} \cite{dakna}.

Concerning the engineering of trapped ionic states, a technique for
preparing even and odd coherent motional states of a trapped ion, via laser
excitation of two vibrational transition, has been described by de Matos
Filho and Vogel \cite{matos}. Starting from the vacuum state as in Ref. \cite%
{matos}, Kneer and Law \cite{kneer} proposed a scheme for generating a
general quantum entanglement of the electronic and vibrational states of a
trapped ion, besides discussing the engineering of two vibrational degrees
of freedom of a single trapped ion. Drobn\'{y} {\it et al.} \cite{buzek}
also provide a technique for deterministic preparation of two-mode
vibrational states. A method for creating deterministic electronic Bell
states of two trapped ions was reported by Solano {\it et al.} \cite{solano}%
, and Moya-Cessa {\it et al}. \cite{mc} presented a procedure to generate
arbitrary discrete superpositions of vibrational coherent states.

The preparation of a variety of nonclassical states has been achieved in
cavity QED \cite{haroche,brune,walther} and trapped-ion phenomena \cite%
{meegato,meemov}. ``Schr\"{o}dinger cat''-like states of both the radiation
field trapped in superconducting cavities \cite{haroche} and quantized
motional states of trapped ions \cite{meegato} have been achieved
experimentally. The preparation of pure photon number states of the
radiation field has been recently reported \cite{walther}.

The feasibility of engineering trapped ionic states relies on the fact that
decoherence of quantum superpositions can be made negligible by suppressing
spontaneous emission using metastable transitions. Coherence of the atomic
population survives for many Rabi cycles of the Jaynes-Cummings (JC)
interaction \cite{wine}, during which, at sufficiently low pressure,
collisions with background atoms can also be avoided \cite{win94}. Here we
mention a recent, remarkable experimental achievement, in which the
decoherence of superposed motional states of a single trapped ion was
controlled through engineered quantum reservoirs \cite{myatt}. To accomplish
this, laser cooling techniques have been considered to generate an
effectively zero-temperature reservoir as suggested in Ref. \cite%
{Poyatos-Marzoli}. Differently from processes in cavity QED, where the
cavity-damping mechanism is the source of errors and decoherence, in the
domain of trapped ions it is assumed that the errors arise from fluctuating
electrical fields of the trap and intensity and phase fluctuations in the
exciting laser pulses \cite{roos}. Stochastic models have been proposed to
deal with this sources of errors \cite{milburn98,milburn99,James,pknight},
and they are shown to be in good qualitative agreement with recent
experiments. Recently, instead of a stochastic mechanism, Di Fidio and Vogel %
\cite{vogeln} proposed a model where the observed damping in Rabi
oscillations \cite{win94} is caused by quantum jumps to an auxiliary
electronic level. Other less important error sources in trapped ions, such
as collisions with the background gas, are also present \cite%
{milburn98,losalamos}{\bf . }

An advantage of engineering trapped ionic states over cavity-field states is
that the latter is a more demanding process. In fact, a bunch of two-level
atoms are required to generate a cavity field, each atom being appropriately
prepared by a Ramsey zone (in a particular superposition state), with its
velocity selected before interacting with the cavity field, and detected
before the passage of the subsequent atom \cite{serra}. On the other hand,
the preparation of an ionic state is designed by just switching laser pulses
on and off, alternately, to manipulate the electronic and vibrational states
of the trapped ion. However, we show here that the errors introduced by the
intensity fluctuation in the exciting laser pulses has a severe effect on
the engineered state, even more dramatic than the errors introduced by the
cavity-damping mechanism when preparing a cavity-field state \cite{norton}.

In this paper we present a scheme for engineering an arbitrary motional
state of a trapped ion by the {\it projection synthesis}{\em \ }method,
which was originally proposed in the travelling-wave domain \cite{pegg} for
the measurement of particular properties of the radiation field, such as its
phase \cite{pegg2} or its $Q$-function \cite{baseia}. This scheme consists
in {\it sculpting} an arbitrary motional ionic state from a coherent
motional state previously prepared in the ionic trap. The technique of
sculpting an arbitrary state from a coherent superposition through the
projection synthesis technique was previously developed in the cavity QED
domain \cite{serra}. However, here we take advantage of the facility to
manipulate trapped ions to make the sculpture process even more attractive
in terms of experimental implementation. In the cavity QED domain, the
sculpture technique \cite{serra} considerably improves previous schemes for
generating an arbitrary state of the radiation field \cite{vogel}. Whereas
in Ref. \cite{vogel} $N$ atoms are required to generate a field with a
maximum number of photons equal to $N$, the sculpture technique requires
about half of this number, exactly $M=int\left[ \left( N+1\right) /2\right] $%
. This is due to the fact that we begin our process from a coherent state
previously injected into the cavity, instead of from the vacuum state as in %
\cite{vogel}. So, instead of building up the desired state photon by photon,
we proceed to sculpt an existing coherent state with an appropriate average
excitation previously calculated. The situation is analogous to the present
proposal for creating motional ionic states. Instead of requiring $N$ steps
to generate an arbitrary motional state with a maximum number of phonons
equal to $N$ \cite{kneer}, our technique utilizes just $M$ steps, as in the
cavity QED context.

We stress that in the present work we elaborate the engineering process of
the vibrational ionic state in the realistic presence of noise. Following
the reasoning in Ref. \cite{milburn98}{\bf ,} we consider the noise arising
from the intensity fluctuations{\bf \ }of the laser pulses used to
manipulate the electronic and vibrational states of the trapped ion. Thus,
after discussing the fundamental interactions between the trapped ion and a
classical field in Sec. II, we describe the sculpture technique for
preparing the ionic vibrational state in the ideal case (absence of noise)
in Sec. III. In Sec. IV, we calculate the fidelity of a sculpted state under
the effects of noise. Here, a phenomenological-operator approach \cite%
{poa,norton} will be applied to account for the presence of noise in the
engineering process. This approach has been developed precisely to compute
the effects of noise in complex processes such as quantum state engineering
and teleportation of quantum states; processes where the protocol requires
several steps of quantum interactions. Through the phenomenological-operator
approach, first developed in the domain of cavity QED \cite{poa,norton}, the
effect of noise is introduced directly in the evolution of the state vector
of the whole system, instead of turning to the evolution of the density
operator as in the {\it ab initio} methods. In Sec. V we present a technique
for optimizing the fidelity of the sculpted state which is based on an
appropriate choice of the ion-laser interaction parameters in the expression
of the sculpted state computed in the presence of noise. Also in Sec. V, we
illustrate the scheme by sculpting a phase state and computing the {\it %
fidelity-probability rate}, a cost-benefit estimate for sculpting a desired
state, as defined in Sec. III. Finally, in Sec. VI we give a summary and
draw conclusions.

\section{Model}

We consider one single trapped ion of mass $m$ in a one-dimensional harmonic
trap whose frequency is $\nu $. The ion has forbidden transitions between two%
{\bf \ }internal electronic states (excited $|\uparrow \rangle $ and ground $%
|\downarrow \rangle $ states, taken as hyperfine sublevels of the ground
state), separated by frequency $\omega _{0}$ and indirectly coupled by
interactions with two laser beams, of frequencies $\omega _{1}$ and $\omega
_{2}$, in a stimulated Raman-type configuration. As indicated in Fig. 1, the
laser beams are detuned by $\Delta $ from a third more excited level $%
|r\rangle $ which, in the stimulated Raman-type configuration, is
adiabatically eliminated when $\Delta $ is much larger than all of the
following: the linewidth of level $|r\rangle $, the coupling associated with
the $|\uparrow \rangle $ $\leftrightarrow $ $|r\rangle $ and $|\downarrow
\rangle $ $\leftrightarrow $ $|r\rangle $ transitions, and the detuning $%
\delta \equiv \omega _{0}-\omega _{L}$ ($\omega _{L}=\omega _{1}-\omega _{2}$%
) \cite{dav,stein}. The transition between $|\downarrow \rangle $ and a
fourth level $\left| d\right\rangle $, achieved by another laser strongly
coupled to the electronic ground state, is considered in order to measure
the ionic vibrational state by collecting the resonance fluorescence signal,
which is the probability of the ion being found in the internal state $%
\left| \downarrow \right\rangle $ \cite{meemov}.

The Hamiltonian that describes the effective interaction of the quantized
motion of the ionic center-of-mass (CM) with its electronic degree of
freedom is \cite{meemov,dav}, in a frame rotating at the ``effective laser
frequency'' $\omega _L$ ( $\hbar =1$): 
\begin{equation}
\widehat{H}=\nu \widehat{a}^{\dagger }\widehat{a}+\frac \delta 2\widehat{%
\sigma }_z+\Omega \left( \widehat{\sigma }_{-}e^{-i\eta (\widehat{a}+%
\widehat{a}^{\dagger })+i\varphi }+\widehat{\sigma }_{+}e^{i\eta (\widehat{a}%
+\widehat{a}^{\dagger })-i\varphi }\right) ,  \label{ham1}
\end{equation}
where $\varphi $ is the phase difference between the two lasers, $\widehat{%
\sigma }_{+}=|\uparrow \rangle \langle \downarrow |$, $\widehat{\sigma }%
_{-}=|\downarrow \rangle \langle \uparrow |$ and $\widehat{\sigma }_z$ are
the usual Pauli pseudo-spin operators, $\widehat{a}^{\dagger }$($\widehat{a}$%
) is the creation (annihilation) operator of vibrational quanta, $\Omega $
is the effective Rabi frequency of the transition $\left| \uparrow
\right\rangle \leftrightarrow \left| \downarrow \right\rangle $ and, $\eta $
is the Lamb-Dicke parameter defined as \cite{meemov,stein}: 
\begin{equation}
\eta =\frac{\Delta k}{\sqrt{2m\nu }}.  \label{eta}
\end{equation}
Here $\Delta k=(\overrightarrow{k}_1-\overrightarrow{k}_2)\cdot 
\overrightarrow{i},|\overrightarrow{k}_{1(2)}|=\omega _{1(2)}/c$, $%
\overrightarrow{k}_1$ ( $\overrightarrow{k}_2$ ) being the wave vector for
laser $1(2)$, and $\overrightarrow{i}$ is the unit vector in the direction
of the trap axis .

Written $H$ in the interaction picture and then expanding the resulting
Hamiltonian in terms of the Lamb-Dicke parameter we get 
\begin{equation}
\widehat{H}_I=\Omega e^{-\eta ^2/2}\left[ \sum_{m,l=0}^\infty \frac{(i\eta
)^{m+l}}{m!l!}e^{-i\left[ (m-l)\nu +\delta \right] t+i\varphi }\widehat{a}^{{%
\dagger }^m}\widehat{a}^l\widehat{\sigma }_{-}+h.c.\right] .  \label{hexp}
\end{equation}
Assuming the Lamb-Dicke limit, for which $\eta \ll 1$, where the ionic CM
motion is strongly localized with respect to the laser wavelengths, we
obtain the simplified Hamiltonian 
\[
\widehat{H}_I=\Omega \left( e^{-i\delta t+i\varphi }\widehat{\sigma }%
_{-}+i\eta e^{-i\left( \nu +\delta \right) t+i\varphi }\widehat{a}^{{\dagger 
}}\widehat{\sigma }_{-}+i\eta e^{-i\left( -\nu +\delta \right) t+i\varphi }%
\widehat{a}\widehat{\sigma }_{-}+h.c.\right) , 
\]
where resonance is achieved by tuning the laser frequencies to obtain $%
\delta =-\ell \nu $ $(\ell =m-l)$. Considering the realistic value for the
trap frequency $\nu /2\pi \approx 11.2$ MHz \cite{wine}, the optical
rotating wave approximation leads to the Carrier Hamiltonian ($\ell =0$)
when tuning the effective laser frequency to obtain $\delta =0$:

\begin{equation}
\widehat{H}_c=\Omega (\widehat{\sigma }_{+}e^{-i\varphi }+\widehat{\sigma }%
_{-}e^{i\varphi }).  \label{car}
\end{equation}
This Hamiltonian induces the transition $\left| n,\downarrow \right\rangle
\longleftrightarrow \left| n,\uparrow \right\rangle $ (where $\left|
n\right\rangle $ indicates a motional Fock state), and is responsible for
rotating only the internal electronic levels of the ion wave function in
accordance with

\begin{mathletters}
\begin{eqnarray}
e^{-i\widehat{H}_c\tau }\left| n,\uparrow \right\rangle &=&\cos \left(
\Omega \tau \right) \left| n,\uparrow \right\rangle -ie^{i\varphi }\sin
\left( \Omega \tau \right) \left| n,\downarrow \right\rangle ,  \label{rota}
\\
e^{-i\widehat{H}_c\tau }\left| n,\downarrow \right\rangle &=&\cos \left(
\Omega \tau \right) \left| n,\downarrow \right\rangle -ie^{-i\varphi }\sin
\left( \Omega \tau \right) \left| n,\uparrow \right\rangle .  \label{rotb}
\end{eqnarray}
When tuning the effective laser frequency to obtain $\delta =-\nu $, the
optical rotating wave approximation leads to the Jaynes-Cummings like
Hamiltonian ($\ell =1$) corresponding to the first red sideband,

\end{mathletters}
\begin{equation}
\widehat{H}_{JC}=i\Omega \eta (\widehat{a}\widehat{\sigma }_{+}e^{-i\varphi
}-\widehat{a}^{{\dagger }}\widehat{\sigma }_{-}e^{i\varphi }),  \label{hjc}
\end{equation}
which induces the transition $\left| n,\downarrow \right\rangle
\longleftrightarrow \left| n-1,\uparrow \right\rangle $, in such a way that
the electronic and vibrational\ modes evolve as

\begin{mathletters}
\begin{eqnarray}
e^{-i\widehat{H}_{JC}\tau }\left| n,\uparrow \right\rangle &=&C_n\left|
n,\uparrow \right\rangle -e^{-i\varphi }S_n\left| n+1,\downarrow
\right\rangle ,  \label{evjca} \\
e^{-i\widehat{H}_{JC}\tau }\left| n,\downarrow \right\rangle
&=&C_{n-1}\left| n,\downarrow \right\rangle +e^{i\varphi }S_{n-1}\left|
n-1,\uparrow \right\rangle ,  \label{evjcb}
\end{eqnarray}
where $C_n=\cos (g\tau \sqrt{n+1})$, $S_n=\sin (g\tau \sqrt{n+1})$, $\tau $
is the duration of the laser pulses, and $g=\Omega \eta $. Finally, we note
that it is possible to obtain the Anti-Jaynes-Cummings Hamiltonian ($\ell
=-1 $) corresponding to the first blue sideband. However, for the purposes
of the present paper we do not use this specific interaction which induces
the transition $\left| n,\downarrow \right\rangle \longleftrightarrow \left|
n+1,\uparrow \right\rangle $.

\section{Sculpture Scheme (ideal case)}

In this section we show how to transpose the {\it projection synthesis}
technique from its original travelling wave domain \cite{pegg} to the
context of the ionic trap (similarly to Ref. \cite{serra} where the {\it %
projection synthesis }was applied to cavity QED phenomena). We assume the
trapped ion to be initially prepared with its CM in a coherent motional
state $\left| \alpha \right\rangle $, and in the electronic excited state $%
\left| \uparrow \right\rangle $. Such a state, which can be prepared with
techniques available nowadays \cite{meegato,meemov}, reads 
\end{mathletters}
\begin{equation}
\left| \Psi ^{(0)}\right\rangle =\left| \alpha \right\rangle \otimes \left|
\uparrow \right\rangle =\sum_{n=0}^\infty \Lambda _n^{(0)}\left|
n\right\rangle \otimes \left| \uparrow \right\rangle ,  \label{psi0}
\end{equation}
with $\Lambda _n^{(0)}=\left. \exp (-\left| \alpha \right| ^2/2)\alpha
^n\right/ \sqrt{n!}$. As the desired state is generated from a previous
coherent state, we have denominated the present scheme a quantum state
sculpture process (as originally done in \cite{serra}). In fact, as shown
below, our strategy consists of modelling the Wigner function of the desired
state, through appropriate laser pulses, from that of the previously
prepared motional coherent state in (\ref{psi0}). The carrier ($C$) and the
Jaynes-Cummings ($JC$) laser pulses work as ``quantum chisels'' on the
initial coherent distribution, as shown in the quantum algorithm notation %
\cite{qalg} depicted in Fig. 2. In the domain of cavity QED, the
quantum-chisels are played by two-level Rydberg atoms which are made to
interact resonantly with a coherent state initially prepared in a high-Q
cavity \cite{serra}.

The whole operation is accomplished in three-step cycles, requiring
successively: (i) a carrier pulse $C_1$, to prepare the electronic state in
a suitable superposition, (ii) a first red sideband pulse $JC$, to entangle
the ionic motional and electronic states, and, finally, (iii) a sequence of
a carrier pulse $C_2$ and a fluorescence measurement of the ionic electronic
state. The third step constitutes the projection synthesis, which enables us
to synthesize the measurement of a particular superposition of the
electronic state and, consequently, to synthesize the projection of the
motional degree of freedom to the desired sculpted state. Thus, the
projection synthesis technique is here applied to both of the entangled
degrees of freedom of the trapped ion, electronic and motional (analogously,
in the cavity QED domain we synthesized the simultaneous projection of the
atomic and the cavity field states \cite{serra}). A fluorescent signal
projects the ion into state $\left| \downarrow \right\rangle $, while the
absence of fluorescence projects it into state $\left| \uparrow
\right\rangle $. For the present purpose the absence of fluorescence is
crucial in preventing the occurrence of recoil of the ionic CM motion. So,
at the end of each cycle the detection of the absence of fluorescence is
required for the successful accomplishment of the engineering process.{\bf \ 
}We emphasize that the duration of the fluorescence measurement is an order
of magnitude smaller than that of the $JC$ pulse and about half that of a $C$
pulse in experiments involving the usual parameter values \cite{wine}. For a 
$JC$ pulse, $g\tau =\left. \pi \right/ 2$ and for a $C$ pulse, $\Omega \tau
=\left. \pi \right/ 2$, the respective durations being around $2\mu s$\ and $%
0.5\mu s$, while the time for a fluorescence signal is around $0.2\mu s$\ %
\cite{salomon}.{\bf \ }We repeat this three-step cycle $M$ times to
synthesize an arbitrary desired state $\left| \Psi _d\right\rangle
=\sum_{n=0}^{N_d}d_n\left| n\right\rangle \otimes \left| \uparrow
\right\rangle $ (see Fig. 2), where it should be noted that the electronic
state $\left| \uparrow \right\rangle $ factorizes. The parameters $M$ and $%
N_d$ (the maximal excitation number of the desired state) are related as
shown below.

Considering the $k$th cycle of the sculpture process, let us start with the
assumption that the ionic state after the ($k-1$)th cycle is

\begin{equation}
\left| \Psi ^{(k-1)}\right\rangle =\sum_{n=0}^\infty \Lambda
_n^{(k-1)}\left| n,\uparrow \right\rangle {.}  \label{psik-1}
\end{equation}
As the first step of the $k$th cycle we prepare, with the carrier pulse $C_1$%
, the electronic state in the superposition ${\cal N}_{\beta _k}\left(
\left| \uparrow \right\rangle +\beta _k\left| \downarrow \right\rangle
\right) $, where ${\cal N}_{\beta _k}=\left( 1+\left| \beta _k\right|
^2\right) ^{-1/2}$ and $\beta _k$ is a complex parameter adjusted by pulse $%
C_1$, fixed at an appropriate duration and phase of the carrier pulse [as in
Eqs. (\ref{rota},\ref{rotb}) ]. In the second step, the first red sideband
pulse $JC$ entangles the ionic motional and electronic states as follows: 
\begin{eqnarray}
\left| \psi ^{(k)}\right\rangle &=&{\cal N}_{\beta _k}\sum_{n=0}^\infty
\Lambda _n^{(k-1)}\left( C_n^{(k)}\left| n,\uparrow \right\rangle
-e^{-i\varphi _k}S_n^{(k)}\left| n+1,\downarrow \right\rangle +\right. 
\nonumber \\
&&\left. +\beta _kC_{n-1}^{(k)}\left| n,\downarrow \right\rangle
+e^{i\varphi _k}\beta _kS_{n-1}^{(k)}\left| n-1,\uparrow \right\rangle
\right) ,  \label{22}
\end{eqnarray}
where $C_m^{(k)}=\cos (g\tau _k\sqrt{m+1})$, $S_m^{(k)}=\sin (g\tau _k\sqrt{%
m+1})$, $\tau _k$ and $\varphi _k$ are the $k$th $JC$ pulse duration and
phase, respectively. Next, in the third step, we have to synthesize the
projection of state (\ref{22}) into a particular electronic superposition
state (of the $k$th cycle) $\left| \chi ^{(k)}\right\rangle ={\cal N}%
_{\varepsilon _k}\left( \left| \uparrow \right\rangle +\varepsilon
_k^{*}\left| \downarrow \right\rangle \right) $, with ${\cal N}_{\varepsilon
_k}=\left( 1+\left| \varepsilon _k\right| ^2\right) ^{-1/2}$. The complex
number $\varepsilon _k$ results from the $k$th rotation of the electronic
states induced by the second carrier pulse $C_2$. As a consequence of the
absence of fluorescent signal (which is the case of a successful
measurement), the ionic state after the projection synthesis is given by 
\begin{equation}
\left| \Psi ^{(k)}\right\rangle ={\cal N}_k\left| \uparrow \right\rangle
\left\langle \chi ^{(k)}|\psi ^{(k)}\right\rangle =\sum_{n=0}^\infty \Lambda
_n^{(k)}\left| n,\uparrow \right\rangle {.}  \label{33}
\end{equation}
The coefficients $\Lambda _n^{(k)}$ result from those in equation (\ref{22})
using the recurrence formula $\Lambda _n^{(k)}={\cal N}_k\Gamma _n^{(k)}$,
where 
\begin{equation}
\Gamma _n^{(k)}=\left( C_n^{(k)}+\varepsilon _k\beta _kC_{n-1}^{(k)}\right)
\Lambda _n^{(k-1)}+e^{i\varphi _k}\beta _kS_n^{(k)}\Lambda
_{n+1}^{(k-1)}-e^{-i\varphi _k}\varepsilon _k\left( 1-\delta _{n,0}\right)
S_{n-1}^{(k)}\Lambda _{n-1}^{(k-1)},  \label{34}
\end{equation}
and the normalization constant ${\cal N}_k$ is given by

\begin{equation}
{\cal N}_k=\left[ \sum_{n=0}^\infty \left| \Gamma _n^{(k)}\right| ^2\right]
^{-1/2}{\rm {.}}  \label{n}
\end{equation}
The pulse $C_2$ and the absence of fluorescence signal (detecting the state $%
\left| \uparrow \right\rangle $) are needed in order to adjust the
measurement of the special superposition $\left| \chi ^{(k)}\right\rangle $,
during which they play the role of electronic state ``polarizers'' and
``analyzers'', respectively - by analogy with light polarization measurement
- and allow us to analyze an arbitrary superposition of $\left| \uparrow
\right\rangle $ and $\left| \downarrow \right\rangle $. Such a measurement %
\cite{freyberger} works as follows: The carrier pulse $C_2$ is appropriately
adjusted so that the electronic state of the ion in the superposition $%
\left| \chi ^{(k)}\right\rangle $ undergoes a unitary transformation to the
state $|\uparrow \rangle $. Otherwise, the electronic state evolves to $%
|\downarrow \rangle $ if it was initially orthogonal to $\left| \chi
^{(k)}\right\rangle $. After the carrier pulse $C_2$ a fluorescence
measurement is required to project $\left| \uparrow \right\rangle $ or $%
\left| \downarrow \right\rangle $. In general, the electronic state after
the $JC$ pulse will be a superposition of these orthogonal states, so that
in unsuccessful cases we find a fluorescence signal (detecting $\left|
\downarrow \right\rangle $), and in successful cases we do not (detecting $%
\left| \uparrow \right\rangle $). Once we find the absence of fluorescence
the electronic state in the $k$th cycle has been projected on the required
superposition $\left| \chi ^{(k)}\right\rangle $. Thus, the projection
synthesis technique is able to measure observables of the form $\left| \chi
^{(k)}\right\rangle \left\langle \chi ^{(k)}\right| $,\ which represent
projection operators with measurable eigenvalues associated with absence
(detecting $\left| \uparrow \right\rangle $) or presence (detecting $\left|
\downarrow \right\rangle $) of a fluorescent signal.

Here we stress an important difference between the present sculpture process
and that used to prepare an arbitrary state in cavity QED \cite{serra}. In
the latter, the projection synthesis is achieved by measuring the required
two-level Rydberg atoms (the quantum chisels for sculpting the field state),
each in a particular superposition state, through a classical field and
ionization chamber detectors. Hence, after the projection synthesis, the
measured state of the $k$th atom is discarded, since it turns to be useless
for the cavity QED process, whereas the electronic state of the ion,
factored as $\left| \uparrow \right\rangle $ in a successful event, is ready
for the next cycle. Therefore, the motional state of the ion is sculpted by
means of its own electronic states.

Since the projection of a particular electronic state in the $M$th cycle is
supposed to finish the sculpture process, the following equality must be
satisfied 
\begin{equation}
{\cal N}_M\left| \uparrow \right\rangle \left\langle \chi ^{(M)}|\psi
^{(M)}\right\rangle =\sum_{n=0}^\infty \Lambda _n^{(M)}\left| n,\uparrow
\right\rangle =\sum_{n=0}^{N_d}d_n\left| n,\uparrow \right\rangle ,
\label{sculpture}
\end{equation}
requiring that $\Lambda _n^M\approx 0$ when $n\geq N_d+1$. This
approximation results in a non-unity {\it fidelity} for the sculpted state.
Usually, the fidelity of a given quantum process is defined to account for
the inevitable errors introduced by the environment due to dissipative
mechanisms \cite{poa,norton}.{\bf \ }However, as mentioned in the
Introduction,{\bf \ }in the domain of trapped ions it is assumed that the
errors arise from noise due to fluctuations in the trap and laser parameters %
\cite{roos}, and to treat this error source we consider the stochastic model
proposed in \cite{milburn98}, where just the intensity fluctuations of the
laser pulses are considered{\bf .} In this section we focus on the ideal
case in which the fidelity is defined to account only for the errors
introduced by the approximation ($\Lambda _n^M\approx 0$ for $n\geq N_d+1$)
inherent in this sculpture scheme. So, this fidelity, which does not account
for the errors introduced by the environment or fluctuations, reads

\begin{equation}
{\cal F}\equiv \left| \left\langle \Psi _{d}|\psi ^{(M)}\right\rangle
\right| ^{2}=\frac{\left| \sum_{n=0}^{N_{d}}d_{n}^{\ast }\Gamma
_{n}^{(M)}\right| ^{2}}{\sum_{l=0}^{\infty }\left| \Gamma _{l}^{(M)}\right|
^{2}}.  \label{fidel}
\end{equation}

The total probability of successfully sculpting the desired state is ${\cal %
P=}\prod_{k=1}^MP_k$, where $P_k$ is the probability of synthesizing a
particular electronic superposition, $\left| \left\langle \chi ^{(k)}|\psi
^{(k)}\right\rangle \right| ^2$, from the $k$th entanglement between
motional and electronic states, Eq. (\ref{22}). In other words, $P_k$ refers
to the probability of measuring absence of fluorescence after the second
carrier pulse in the $k$th cycle. This probability is given by

\begin{equation}
P_{k}=\left| \left\langle \chi ^{(k)}|\psi ^{(k)}\right\rangle \right| ^{2}=%
{\cal N}_{\varepsilon _{k}}^{2}{\cal N}_{\beta _{k}}^{2}\sum_{n=0}^{\infty
}\left| \Gamma _{n}^{k}\right| ^{2}.  \label{prob}
\end{equation}

For an appropriate choice of average excitation of the coherent motional
state, $\left| \alpha \right| ^{2}=\overline{n}_{\alpha }$, we see from the
recurrence formula (\ref{34}) and the definition of coefficients $\Lambda
_{n}^{(0)}$, that the coefficients $\Gamma _{n}^{(M)}$ depend on powers of $%
\alpha $, varying as $\alpha ^{j}/\sqrt{j!}$, with $n-M\leq j\leq n+M$. In
fact, it is straightforward to conclude that, after one application of
formula (\ref{34}) the coefficients $\Gamma _{n}^{(M)}$ are proportional to $%
\left\{ \Lambda _{n-1}^{(M-1)},\Lambda _{n}^{(M-1)},\Lambda
_{n+1}^{(M-1)}\right\} $; after two applications of the formula (\ref{34})
it follows that $\Gamma _{n}^{(M)}\propto \left\{ \Lambda
_{n-2}^{(M-2)},\Lambda _{n-1}^{(M-2)},\Lambda _{n}^{(M-2)},\Lambda
_{n+1}^{(M-2)},\Lambda _{n+2}^{(M-2)}\right\} $ and after $M$ applications
we finally obtain $\Gamma _{n}^{(M)}\propto \left\{ \Lambda
_{n-M}^{(0)},\Lambda _{n-M+1}^{(0)},...,\Lambda _{n}^{(0)},...,\Lambda
_{n+M-1}^{(0)},\Lambda _{n+M}^{(0)}\right\} $. Therefore, from the
definition of $\Lambda _{n}^{(0)}$ (initial coherent motional state), we
note that $\Gamma _{n}^{(M)}$ depends on powers of $\alpha $, as mentioned
above, and the choice of the average excitation $\overline{n}_{\alpha }$
that ensures $P(N_{d}-M+1)=\left| \left\langle N_{d}-M+1|\alpha
\right\rangle \right| ^{2}\approx 0$ (satisfying the requirement that $%
\Lambda _{n}^{(M)}\approx 0$ when $n\geq N_{d}+1$) results in a higher
fidelity ${\cal F}$ at the expense of a lower probability ${\cal P}$. In
fact, it is evident from the denominator of Eq. (\ref{fidel}) that the lower
the number $l$ in the sum of significant coefficients, the higher the
fidelity. On the other hand, in Eq. (\ref{prob}) we observe that the
probability ${\cal P}$ is directly proportional to powers of $\alpha $. As a
consequence, Eqs. (\ref{fidel}) and (\ref{prob}) furnish a
fidelity-probability rate, ${\cal R}\equiv {\cal F}^{\xi }{\cal P}^{\zeta }$%
, a cost-benefit estimate for sculpting the desired state, where the
parameters $\xi $ and $\zeta $ are appropriately chosen to weight the
contributions of the fidelity and the probability in accordance with the
aims of the sculptor. In fact, the sculptor may decide to privilege the
fidelity or the probability in the cost-benefit estimative, and in the
present work we have decided to privilege the fidelity, choosing the values $%
\xi =4$ and{\bf \ }$\zeta =1/2$. In order to maximize the rate ${\cal R}$,
we have to play with all the parameters: the durations and phases of the $JC$
laser pulses, $\tau _{k}$ and $\varphi _{k}$, and the carrier pulse ($C_{1}$
and $C_{2}$) parameters $\beta _{k}$ and $\varepsilon _{k}$. We note that a
good strategy to maximize ${\cal R}$ consists in starting with a choice of $%
\overline{n}_{\alpha }$ so that $P(N_{d}-M+1)\approx 0$, and then proceed to
maximize the rate ${\cal R}$, increasing $\overline{n}_{\alpha }$ at the
expense of the fidelity. Next, the duration and phase of the $JC$ laser
pulses $\tau _{k}$ and $\varphi _{k}$ are chosen so as to maximize the rate $%
{\cal R}$. Finally, the choice of the carrier pulse parameters, $\beta _{k}$
and $\varepsilon _{k}$, follows from a particular solution of the equality (%
\ref{sculpture}) which, togheter with the requirement $\Lambda _{n}^{\left(
M\right) }\approx 0$ when $n\geq N_{d}+1$, results in the set of $N_{d}+1$
equations

\begin{eqnarray}
d_{N_{d}} &=&{\cal N}_{M}\left[ \left( C_{N_{d}}^{(M)}+\varepsilon _{M}\beta
_{M}C_{N_{d}-1}^{(M)}\right) \Lambda _{N_{d}}^{(M-1)}+e^{i\varphi _{k}}\beta
_{M}S_{N_{d}}^{(M)}\Lambda _{N_{d}+1}^{(M-1)}-e^{-i\varphi _{k}}\varepsilon
_{M}S_{N_{d}-1}^{(M)}\Lambda _{N_{d}-1}^{(M-1)}\right] ,  \nonumber \\
\vdots &=&\vdots  \nonumber \\
d_{n} &=&{\cal N}_{M}\left[ \left( C_{n}^{(M)}+\varepsilon _{M}\beta
_{M}C_{n-1}^{(M)}\right) \Lambda _{n}^{(M-1)}+e^{i\varphi _{k}}\beta
_{M}S_{n}^{(M)}\Lambda _{n+1}^{(M-1)}-e^{-i\varphi _{k}}\varepsilon
_{M}S_{n-1}^{(M)}\Lambda _{n-1}^{(M-1)}\right] ,  \label{7a} \\
\vdots &=&\vdots  \nonumber \\
d_{0} &=&{\cal N}_{M}\left[ \left( C_{0}^{(M)}+\varepsilon _{M}\beta
_{M}\right) \Lambda _{0}^{(M-1)}+e^{i\varphi _{k}}\beta
_{M}S_{0}^{(M)}\Lambda _{1}^{(M-1)}\right] .  \nonumber
\end{eqnarray}

To solve the set of equations (\ref{7a}), we apply the recurrence formula (%
\ref{34}) $M-1$ times in order to express the unknown coefficients $\Lambda
_n^{(k)}$ in terms of the known values of the coefficients of the coherent
motional state, $\Lambda _n^{(0)}$. In this way we obtain a nonlinear system
whose free parameters are $\beta _1,...,\beta _M$ and $\varepsilon
_1,...,\varepsilon _M$, which indicate the particular rotation the
electronic state must undergo in each cycle, during each carrier pulse ($C_1$
and $C_2$). These variables are obtained from the known coefficients $d_n$
and $\Lambda _n^{\left( 0\right) }$. The solvability of a nonlinear system
can be ensured if the number of equations is equal to the number of
variables, the latter being the parameters of both carrier pulses. One of
the equations in system (\ref{7a}) has to be used to obtain the
normalization constant $\prod_{k=1}^M{\cal N}_k$ and each cycle carries two
free parameters ($\beta _k,\varepsilon _k$). Therefore, the minimum number
of cycles necessary to guarantee the solution of system (\ref{7a}) must be $%
M=$int$\left[ \left( N_d+1\right) /2\right] $. This conclusion follows from
the fact that in our scheme we start from a coherent motional state. The
real variables $\tau _1,\tau _2,...,\tau _M$\ and $\phi _1,\phi _2...\phi _M$%
\ are used to improve the cost-benefit rate$\ R$.

With this technique, therefore, it is possible to sculpt an arbitrary
vibrational state by remodelling another initial vibrational state. In
particular, we have started from the coherent state since it is easily
generated\cite{meemov}.

\subsection{Sculpting a truncated phase state}

To illustrate the sculpture technique we now proceed to engineer the
truncated phase state ($N_d=2$) 
\begin{equation}
\left| \Psi _d\right\rangle =\frac 1{\sqrt{3}}\sum_{n=0}^2\left| n,\uparrow
\right\rangle .  \label{desejado}
\end{equation}
As mentioned above, the sculpture process for the state $\left| \Psi
_d\right\rangle $ requires just $M=1$ cycle, and we obtain from equations (%
\ref{7a}) the system

\begin{mathletters}
\begin{eqnarray}
\frac{\left[ \left( C_2^{(1)}+\varepsilon _1\beta _1C_1^{(1)}\right) \Lambda
_2^{(0)}+e^{i\varphi _1}\beta _1S_2^{(1)}\Lambda _3^{(0)}-e^{-i\varphi
_1}\varepsilon _1S_1^{(0)}\Lambda _1^{(0)}\right] }{\left[ \left(
C_0^{(1)}+\varepsilon _1\beta _1\right) \Lambda _0^{(0)}+e^{i\varphi
_1}\beta _1S_0^{(1)}\Lambda _1^{(0)}\right] } &=&1  \label{s2a} \\
\frac{\left[ \left( C_1^{(M)}+\varepsilon _1\beta _1C_0^{(1)}\right) \Lambda
_1^{(0)}+e^{i\phi _1}\beta _1S_1^{(1)}\Lambda _2^{(0)}-e^{-i\phi
_1}\varepsilon _1S_0^{(1)}\Lambda _0^{(0)}\right] }{\left[ \left(
C_0^{(1)}+\varepsilon _1\beta _1\right) \Lambda _0^{(0)}+e^{i\varphi
_1}\beta _1S_0^{(1)}\Lambda _1^{(0)}\right] } &=&1  \label{s2b}
\end{eqnarray}
Solving the above system we obtain a fourth-order polynomial equation in the
variable $\varepsilon _1$ ($\beta _1$) by isolating the variable $\beta _1$ (%
$\varepsilon _1$) from one equation and substituting it into the other. In
this way we obtain the roots of the system (\ref{s2a},\ref{s2b}) for any
fixed set of parameters $\overline{n}_\alpha $, $g\tau _1$\ and $\varphi _1$%
. When considering more than one cycle to sculpt a state where $N_d>2$,
instead of Eqs. (\ref{s2a},\ref{s2b}) we obtain a set of $N_d$ coupled
equations, permitting only numerical solutions (\cite{serra}).

Following the strategy mentioned above, we start with the average excitation 
$\overline{n}_\alpha $ leading to the highest fidelity resulting from $%
P(3)=\left| \left\langle 2|\alpha \right\rangle \right| ^2\approx 0$. For
our purposes we begin with the average excitation $\overline{n}_\alpha =0.04$%
. For each value of $\overline{n}_\alpha $\ (choosing $\alpha $\ as a real
parameter) we proceed to calculate the duration ($g\tau _1$) and phase ($%
\varphi _1$) of the $JC$\ laser pulse which maximize the rate ${\cal R}$. As
discussed above, the maximum value of ${\cal R}$\ depends on the choice of
the parameters $\xi =4$ and $\zeta =1/2$, weighting ${\cal F}$ and ${\cal P}$%
, respectively. In Table I we show the rate ${\cal R}$\ associated with each
value of $\overline{n}_\alpha $, from that which maximizes the fidelity ($%
0.04$) to values that exhibit a continuous decrease of the rate ${\cal R}$.
In addition, four roots ($\varepsilon _1$,$\beta _1$) of Eqs. (\ref{s2a},\ref%
{s2b}) result when a given pair of parameters ($g\tau _1$,$\varphi _1$) are
fixed, and we have to choose the one which maximizes ${\cal R}$. The value $%
\overline{n}_\alpha =0.25$\ results in the highest rate ${\cal R}=0.60$,
which follows from a fidelity ${\cal F}=0.99$\ and probability ${\cal P}%
=0.38 $ of successfully sculpting the desired state.

We display in Figs. 3(a),3(b)\ the sculpture process of the desired
truncated phase state (\ref{desejado}) from the Wigner distribution function
of the initial coherent state (associated to $\overline{n}_\alpha =0.25$)
given by a Gaussian shifted from the origin as $W(p,q)=$\ $\left( 2/\pi
\right) $\ $\exp \left[ -\left( q+\alpha \right) ^2-p^2\right] $, shown in
Fig. 3(a). The state (associated with the best rate ${\cal R}=0.60$)
obtained after one cycle is displayed in Fig.3(b), using the parameters ($%
g\tau _1$,$\varphi _1,\varepsilon _1$\ and $\beta _1$) associated with this
rate. As we have stressed above, it is possible by the present scheme to
sculpt the desired state with a higher probability of sucess but at the
expense of a smaller fidelity. So, the sculpture technique can be evaluated
by a cost-benefit estimate, here defined as the fidelity-probability rate,
which it is up to the sculptor to choose.

The next section deals with the effects of noise on the sculpture process,
i.e., the influence of the errors arising from the intensity fluctuations in
the exciting laser pulses. In this paper we do not consider either the
(weaker) effects of the phase fluctuations of the laser pulses \cite%
{milburn98} or the (close to unity) efficiency of detection of the internal
state required in the third step of the cycle \cite{roos}. When the effects
of noise on the sculpture process are taken into account, following the
phenomenological-operator approach, the fidelity defined by Eq. (\ref{fidel}%
) remains exactly the same, whereas the sculpt field state, after the last
required cycle, will be entangled with auxiliary states in which
noise-operators are defined. So, to the best of our knowledge the sculpted
motional state will be represented by a statistical mixture $\rho _{ion}$,
whereby the fidelity turns out to be ${\cal F}=\left\langle \psi _d\left|
\rho _{ion}\right| \psi _d\right\rangle $.

\section{Effects of noise on the process}

As mentioned above, we consider here the noise arising from the intensity
fluctuations in the exciting laser pulses (carrier and Jaynes-Cummings)
required to manipulate the internal and external states of the trapped ion %
\cite{milburn98,losalamos}. We will also propose an alternative way of
engineering an ionic motional state in the presence of noise, which consists
in maximizing the fidelity of the experimentally achieved state. A
phenomenological-operator approach, originally developed in the context of
cavity QED \cite{poa,norton}, will be considered here to account for the
evolution of the ionic states under the influence of the
fluctuating-intensity laser pulses{\bf . }The strategy is to provide a
straightforward technique to incorporate the main results obtained by
standard ab-initio methods for treating errors in trapped ions (in
particular we consider the master equation calculations in Ref. \cite%
{milburn98}). In short, we introduce an auxiliary state space where noise
operators are defined to allow for the effects of noise explicitly in the
evolution of the state vector of the whole system comprehended by the ionic
and auxiliary states{\bf . }After computing the evolved state of the whole
system, the reduced density matrix of the ionic system can immediately be
obtained by tracing out the auxiliary variables. The phenomenological
approach is constructed so that the reduced density matrix turns out to be
exactly the same as the one obtained by standard methods. It is interesting
to note that the phenomenological approach resembles the Monte Carlo wave
function method \cite{mcwf} in the sense that we work directly with the wave
function, providing an efficient computational tool.

First we give a short description of the master equation treatment by
Schneider and Milburn \cite{milburn98} of the laser fluctuations as a
stochastic process, the rise or fall of the laser intensity being defined as
a real Wiener process. After the pulses required to generate the desired
state, the noise introduced is effectively averaged and the master equation
describing the ionic system follows from the stochastic Liouville-von
Neumann equation \cite{milburn98}

\end{mathletters}
\begin{equation}
\frac d{dt}\widehat{\rho }\left( t\right) =-i\left[ \widehat{{\cal H}},%
\widehat{\rho }\left( t\right) \right] -\frac \Gamma 2\left[ \widehat{{\cal H%
}},\left[ \widehat{{\cal H}},\widehat{\rho }\left( t\right) \right] \right]
\label{noise}
\end{equation}
where $\widehat{{\cal H}}$ is the interaction Hamiltonian for the $C$ (\ref%
{car}) or $JC$ (\ref{hjc}) pulse and the parameter $\Gamma $, to be obtained
phenomenologically, scales the noise. Considering the subspace composed of
the eigenstates of $\widehat{{\cal H}}$, it is straightforward to solve Eq. (%
\ref{noise}), obtaining

\begin{eqnarray}
\left\langle \Phi _{n}^{\pm }\right| \rho \left( t\right) \left| \Phi
_{m}^{\pm }\right\rangle &=&\exp \left[ -it\left( \Phi _{n}^{\pm }-\Phi
_{m}^{\pm }\right) -2\Gamma t\left( \Phi _{n}^{\pm }-\Phi _{m}^{\pm }\right)
^{2}\right] \times  \nonumber \\
&&\times \left\langle \Phi _{n}^{\pm }\right| \rho \left( 0\right) \left|
\Phi _{m}^{\pm }\right\rangle  \label{elem}
\end{eqnarray}
where $\left| \Phi _{m}^{\pm }\right\rangle $ are the eigenstates and $\Phi
_{n}^{\pm }$ are the eigenvalues for the interaction Hamiltonian described
by Eqs. (\ref{car}) and (\ref{hjc}). For the $C$ Hamiltonian,

\begin{eqnarray*}
\left| \Phi _n^{\pm }\right\rangle &=&\frac 1{\sqrt{2}}\left( \left|
n\downarrow \right\rangle \pm e^{i\varphi }\left| n\uparrow \right\rangle
\right) , \\
\Phi _n^{\pm } &=&\pm \Omega ,
\end{eqnarray*}
while for the $JC$ Hamiltonian,

\begin{eqnarray*}
\left| \Phi _{n}^{\pm }\right\rangle &=&\frac{1}{\sqrt{2}}\left( \left|
n,\downarrow \right\rangle \pm ie^{i\phi }\left| n-1,\uparrow \right\rangle
\right) , \\
\Phi _{n}^{\pm } &=&\pm g\sqrt{n}.
\end{eqnarray*}

Employing the reasoning of the phenomenological-operator approach, we have
introduced the noise-operators $\widehat{{\cal C}}$ and $\widehat{{\cal J}}$%
, to account for the intensity fluctuations in the $C$ and the $JC$ pulses,
respectively. These operators are supposed to act on the auxiliary states $%
\left| {\bf C}\right\rangle $ and $\left| {\bf J}\right\rangle $, so as to
the effects of noise explicitly in the evolution of the whole system, now
composed of the ionic and auxiliary states. After the ion-laser interactions
required for the sculpture process, the reduced density operator of the ion
is obtained by tracing out the auxiliary space describing the noise-sources.

With the noise-operators defined in the auxiliary spaces, we observe that
the coupling of a general ionic state to the fluctuating $C$ laser pulses
evolves in time as follows: 
\begin{eqnarray}
\sum_n\left( \alpha _n\left| n\downarrow \right\rangle +\beta _n\left|
n\uparrow \right\rangle \right) \otimes \left| {\bf C}\right\rangle
&\longrightarrow &\left| \Psi _C\right\rangle =\sum_n\left( \alpha _n%
\widehat{{\cal C}}_{n,\downarrow \downarrow }\left( t,\varphi \right) +\beta
_n\widehat{{\cal C}}_{n,\uparrow \downarrow }\left( t,\varphi \right)
\right) \left| n\downarrow \right\rangle \otimes \left| {\bf C}\right\rangle
+  \nonumber \\
&&\left( \alpha _n\widehat{{\cal C}}_{n,\downarrow \uparrow }\left(
t,\varphi \right) +\beta _n\widehat{{\cal C}}_{n,\uparrow \uparrow }\left(
t,\varphi \right) \right) \left| n\uparrow \right\rangle \otimes \left| {\bf %
C}\right\rangle  \label{evc}
\end{eqnarray}
while for the $JC$ pulse it follows

\begin{eqnarray}
\sum_n\left( \alpha _n\left| n,\downarrow \right\rangle +\beta _n\left|
n,\uparrow \right\rangle \right) \otimes \left| {\bf J}\right\rangle
&\longrightarrow &\left| \Psi _{JC}\right\rangle =\sum_n\left[ \alpha _n%
\widehat{{\cal J}}_{n,\downarrow \downarrow }\left( t,\varphi \right) \left|
n,\downarrow \right\rangle +\beta _n\widehat{{\cal J}}_{n,\uparrow
\downarrow }\left( t,\varphi \right) \left| n+1,\downarrow \right\rangle
+\right.  \nonumber \\
&&\left. \alpha _n\widehat{{\cal J}}_{n\downarrow \uparrow }\left( t,\varphi
\right) \left| n-1,\uparrow \right\rangle +\beta _n\widehat{{\cal J}}%
_{n,\uparrow \uparrow }\left( t,\varphi \right) \left| n,\uparrow
\right\rangle \right] \otimes \left| {\bf J}\right\rangle {\rm {.}}
\label{evjc}
\end{eqnarray}

Following the reasoning of the phenomenological approach, the matrix elements%
{\bf \ }$\left\langle {\bf C}\left| \widehat{{\cal C}}_{n,jk}\left(
t,\varphi \right) \widehat{{\cal C}}_{n^{\shortmid }j^{\shortmid
}k^{\shortmid }}^{\dagger }\left( t,\varphi \right) \right| {\bf C}%
\right\rangle ${\bf \ }and $\left\langle {\bf J}\left| \widehat{{\cal J}}%
_{n,jk}\left( t,\varphi \right) \widehat{{\cal J}}_{n^{\shortmid
},j^{\shortmid }k^{\shortmid }}^{\dagger }\left( t,\varphi \right) \right| 
{\bf J}\right\rangle $\ ($n,n^{\shortmid }=0,1,2....$, $j,$\ $j^{\shortmid
}=\uparrow ,\downarrow $\ and $k,k^{\shortmid }=\uparrow ,\downarrow $),
which result from tracing the density operator associated with Eqs. (\ref%
{evc}) and (\ref{evjc})\ over the auxiliary spaces\ ${\bf C}$\ and ${\bf J}$
(Tr$_{{\bf C}({\bf JC})}$ $\left| \Psi _{C(JC)}\right\rangle \left\langle
\Psi _{C(JC)}\right| =\rho _{C(JC)}^{red}$), are inferred from the standard
stochastic method in \cite{milburn98}. Comparing the reduced density
operator $\rho _{C(JC)}^{red}$ with those emerging from the evolution in Eq.
(\ref{elem}) of the density matrix associated with the state\ $\sum_n\left(
\alpha _n\left| n\downarrow \right\rangle +\beta _n\left| n\uparrow
\right\rangle \right) $, we obtain the required matrix elements shown in
Appendix A. Such matrix elements can now be directly applied to any process
that involves interactions of an ion with fluctuating intensity laser
pulses, removing the necessity to perform the typically extensive {\it ab
initio} calculations. In the case of a process requiring several laser
pulses (as when sculpting a state with a large $N_d$) it turns to be
practically prohibitive to compute the evolution of the ionic system by
standard techniques.

\subsection{Estimating the errors introduced by $C$ and $JC$ pulses}

Let us now consider the initial state $\left| \psi _C\left( 0\right)
\right\rangle =\left| \psi _{JC}\left( 0\right) \right\rangle =\left|
n,\downarrow \right\rangle $, which evolves, in the ideal case (without
fluctuations), following Eqs. (\ref{rotb}) and (\ref{evjcb}) for carrier and
Jaynes-Cummings pulses, respectively:

\begin{eqnarray*}
\left| \psi _C\left( t\right) \right\rangle &=&e^{-i\widehat{H}_c\tau
}\left| \psi _c\left( 0\right) \right\rangle =\cos \left( \Omega t\right)
\left| n,\downarrow \right\rangle -ie^{-i\varphi }\sin \left( \Omega
t\right) \left| n,\uparrow \right\rangle , \\
\left| \psi _{JC}\left( t\right) \right\rangle &=&e^{-i\widehat{H}_{JC}\tau
}\left| \psi _{JC}\left( 0\right) \right\rangle =\cos \left( gt\sqrt{n}%
\right) \left| n,\downarrow \right\rangle +e^{i\varphi }\sin \left( gt\sqrt{n%
}\right) \left| n-1,\uparrow \right\rangle ,
\end{eqnarray*}
where $t$ and $\varphi $ are the duration and phase for both pulses. Now,
given realistic intensity fluctuations in the laser pulses, the evolution of
states $\left| \psi _c\left( 0\right) \right\rangle $ and $\left| \psi
_{JC}\left( 0\right) \right\rangle $ is easily computed with the
phenomenological approach, leading to the results

\begin{eqnarray*}
\left| \widetilde{\psi }_{C}\left( t\right) \right\rangle \otimes \left| 
{\bf C}\right\rangle &=&\left[ \widehat{{\cal C}}_{n,\downarrow \downarrow
}\left( t,\varphi \right) \left| n,\downarrow \right\rangle +\widehat{{\cal C%
}}_{n,\downarrow \uparrow }\left( t,\varphi \right) \left| n,\uparrow
\right\rangle \right] \otimes \left| {\bf C}\right\rangle , \\
\left| \widetilde{\psi }_{JC}\left( t\right) \right\rangle \otimes \left| 
{\bf J}\right\rangle &=&\left[ \widehat{{\cal J}}_{n,\downarrow \downarrow
}\left( t,\varphi \right) \left| n,\downarrow \right\rangle +\widehat{{\cal J%
}}_{n,\downarrow \uparrow }\left( t,\varphi \right) \left| n-1,\uparrow
\right\rangle \right] \otimes \left| {\bf J}\right\rangle .
\end{eqnarray*}
The reduced density operator of the ion (obtained by tracing out the
auxiliary spaces ${\bf C}$ and ${\bf J}$) are

\begin{eqnarray*}
\widehat{\widetilde{\rho }}_C\left( t\right) &=&{\rm {Tr}}_{{\bf C}}\left| 
\widetilde{\psi }_C\left( t\right) \right\rangle \left| {\bf C}\right\rangle
\left\langle {\bf C}\right| \left\langle \widetilde{\psi }_C\left( t\right)
\right| \\
&=&\left\langle \widehat{{\cal C}}_{n,\downarrow \downarrow }\left(
t,\varphi \right) \widehat{{\cal C}}_{n,\downarrow \downarrow }^{\dagger
}\left( t,\varphi \right) \right\rangle _{{\bf C}}\left| n,\downarrow
\right\rangle \left\langle n,\downarrow \right| +\left\langle \widehat{{\cal %
C}}_{n,\downarrow \downarrow }\left( t,\varphi \right) \widehat{{\cal C}}%
_{n,\downarrow \uparrow }^{\dagger }\left( t,\varphi \right) \right\rangle _{%
{\bf C}}\left| n,\downarrow \right\rangle \left\langle n,\uparrow \right| +
\\
&&\left\langle \widehat{{\cal C}}_{n,\downarrow \uparrow }\left( t,\varphi
\right) \widehat{{\cal C}}_{n,\downarrow \downarrow }^{\dagger }\left(
t,\varphi \right) \right\rangle _{{\bf C}}\left| n,\uparrow \right\rangle
\left\langle n,\downarrow \right| +\left\langle \widehat{{\cal C}}%
_{n,\downarrow \uparrow }\left( t,\varphi \right) \widehat{{\cal C}}%
_{n,\downarrow \uparrow }^{\dagger }\left( t,\varphi \right) \right\rangle _{%
{\bf C}}\left| n,\uparrow \right\rangle \left\langle n,\uparrow \right| ,
\end{eqnarray*}

\begin{eqnarray*}
\widehat{\widetilde{\rho }}_{JC}\left( t\right) &=&{\rm {Tr}}_{{\bf J}%
}\left| \widetilde{\psi }_{JC}\left( t\right) \right\rangle \left| {\bf J}%
\right\rangle \left\langle {\bf J}\right| \left\langle \widetilde{\psi }%
_c\left( t\right) \right| \\
&=&\left\langle \widehat{{\cal J}}_{n,\downarrow \downarrow }\left(
t,\varphi \right) \widehat{{\cal J}}_{n,\downarrow \downarrow }^{\dagger
}\left( t,\varphi \right) \right\rangle _{{\bf J}}\left| n,\downarrow
\right\rangle \left\langle n,\downarrow \right| +\left\langle \widehat{{\cal %
J}}_{n,\downarrow \downarrow }\left( t,\varphi \right) \widehat{{\cal J}}%
_{n,\downarrow \uparrow }^{\dagger }\left( t,\varphi \right) \right\rangle _{%
{\bf J}}\left| n,\downarrow \right\rangle \left\langle n-1,\uparrow \right| +
\\
&&\left\langle \widehat{{\cal J}}_{n,\downarrow \uparrow }\left( t,\varphi
\right) \widehat{{\cal J}}_{n,\downarrow \downarrow }^{\dagger }\left(
t,\varphi \right) \right\rangle _{{\bf J}}\left| n-1,\uparrow \right\rangle
\left\langle n,\downarrow \right| +\left\langle \widehat{{\cal J}}%
_{n,\downarrow \uparrow }\left( t,\varphi \right) \widehat{{\cal J}}%
_{n,\downarrow \uparrow }^{\dagger }\left( t,\varphi \right) \right\rangle _{%
{\bf J}}\left| n-1,\uparrow \right\rangle \left\langle n-1,\uparrow \right| .
\end{eqnarray*}

From the matrix elements shown in Appendix A and after a straightforward
calculation we obtain the following fidelities of the evolved states $\left| 
\widetilde{\psi }_C\left( t\right) \right\rangle $ and $\left| \widetilde{%
\psi }_{JC}\left( t\right) \right\rangle $ with respect to the ideal
evolution of $\left| \psi _C\left( 0\right) \right\rangle $ and $\left| \psi
_{JC}\left( 0\right) \right\rangle $

\[
{\cal F}_{C}=\left\langle \psi _{C}\left( t\right) \left| \widehat{%
\widetilde{\rho }}_{c}\left( t\right) \right| \psi _{C}\left( t\right)
\right\rangle =\frac{1}{2}+\frac{1}{2}e^{-2\Gamma \Omega ^{2}t},
\]

\[
{\cal F}_{JC}=\left\langle \psi _{JC}\left( t\right) \left| \widehat{%
\widetilde{\rho }}_{JC}\left( t\right) \right| \psi _{JC}\left( t\right)
\right\rangle =\frac{1}{2}+\frac{1}{2}e^{-2n\Gamma g^{2}t},
\]%
for carrier and Jaynes-Cummings pulses, respectively. Note that for $\Gamma
=0$ we have ${\cal F}_{c}={\cal F}_{JC}=1$, as expected. Besides, for $n=0$
we have ${\cal F}_{JC}=1$ since the intensity fluctuations in the $JC$ pulse
do not exchange energy with the ionic system, maintaining the initial state $%
\left| n=0,\downarrow \right\rangle $ unaffected. As defined in Sec. II, $%
g=\eta \Omega $, and the Lamb-Dicke parameter used in experiments is $\eta
=0.202$ \cite{wine}; thus, the noise introduced by the $JC$ pulse has less
effect than that coming from the $C$ pulse for $n$ $\lesssim 25$. Thus, when
sculpting the phase state (\ref{desejado}) or in any process involving a
small number of phonons, such as engineering a motional qubit state $%
c_{0}\left| 0\right\rangle +c_{1}\left| 1\right\rangle $ for quantum
computation, as performed in \cite{CNOT}, the main source of errors will
undoubtedly be the $C$ pulses.

\subsection{Sculpting a truncated phase state in the presence of noise}

Now we proceed to incorporate the intensity fluctuations of the laser pulses
used in sculpting the truncated phase state (\ref{desejado}) which is
obtained after one cycle of the sculpture process. We assume, for
simplicity, that there is no noise in the preparation of the initial
coherent motional state of the ion. After the preparation of such a motional
state and the first carrier pulse it follows that:

\begin{eqnarray*}
\sum_{n=0}^\infty \Lambda _n^{(0)}\left| n,\uparrow \right\rangle \otimes
\left| {\bf C}_1\right\rangle &\longrightarrow &\sum_{n=0}^\infty \Lambda
_n^{(0)}\left( \widehat{{\cal C}}_{n,\uparrow \downarrow }\left( t_1\right)
\left| n\downarrow \right\rangle +\widehat{{\cal C}}_{n,\uparrow \uparrow
}\left( t_1\right) \left| n\uparrow \right\rangle \right) \otimes \left| 
{\bf C}_1\right\rangle \\
&=&\left| \psi ^{^I}\right\rangle ,
\end{eqnarray*}
where $t_1$ is the duration of the first carrier pulse and ${\bf C}_1$
stands for its auxiliary space. To simplify the notation we have omitted
that the operator $\widehat{{\cal C}}$ depends on the phase $\varphi _1$ of
the $C$ laser pulse. Note that in this first step the required superposition
of the electronic state ${\cal N}_\beta \left( \left| \uparrow \right\rangle
+\beta _k\left| \downarrow \right\rangle \right) $ is obtained with a
non-unity fidelity due to laser fluctuations. In the second step, the first
red sideband pulse $JC$ entangles the ionic motional and electronic states
as follows:

\begin{eqnarray*}
\left| \psi ^{^I}\right\rangle \otimes \left| {\bf J}\right\rangle
&\longrightarrow &\sum_{n=0}^\infty \Lambda _n^{(0)}\left[ \widehat{{\cal J}}%
_{n\downarrow \downarrow }\left( t_2\right) \widehat{{\cal C}}_{n\uparrow
\downarrow }\left( t_1\right) \left| n,\downarrow \right\rangle +\widehat{%
{\cal J}}_{n\uparrow \downarrow }\left( t_2\right) \widehat{{\cal C}}%
_{n\uparrow \uparrow }\left( t_1\right) \left| n+1,\downarrow \right\rangle
+\right. \\
&&\left. +\widehat{{\cal J}}_{n\downarrow \uparrow }\left( t_2\right) 
\widehat{{\cal C}}_{n\uparrow \downarrow }\left( t_1\right) \left|
n-1,\uparrow \right\rangle +\widehat{{\cal J}}_{n\uparrow \uparrow }\left(
t_2\right) \widehat{{\cal C}}_{n\uparrow \uparrow }\left( t_1\right) \left|
n,\uparrow \right\rangle \right] \otimes \left| {\bf J},{\bf C}%
_1\right\rangle \\
&=&\left| \psi ^{^{II}}\right\rangle ,
\end{eqnarray*}
where $t_2$ and $\varphi _2$ are the duration and phase of the $JC$ laser
pulse. Finally, after the second $C$ pulse we have

\begin{eqnarray*}
\left| \psi ^{^{II}}\right\rangle \otimes \left| {\bf C}_2\right\rangle
&\longrightarrow &\sum_{n=0}^\infty \Lambda _n^{(0)}\left[ \left( \widehat{%
{\cal C}}_{n,\downarrow \downarrow }\left( t_3\right) \widehat{{\cal J}}%
_{n,\downarrow \downarrow }\left( t_2\right) \widehat{{\cal C}}_{n,\uparrow
\downarrow }\left( t_1,\varphi _1\right) +\widehat{{\cal C}}_{n,\uparrow
\downarrow }\left( t_3\right) \widehat{{\cal J}}_{n,\uparrow \uparrow
}\left( t_2\right) \widehat{{\cal C}}_{n,\uparrow \uparrow }\left(
t_1\right) \right) \left| n,\downarrow \right\rangle +\right. \\
&&+\widehat{{\cal C}}_{n+1,\downarrow \downarrow }\left( t_3\right) \widehat{%
{\cal J}}_{n\uparrow \downarrow }\left( t_2\right) \widehat{{\cal C}}%
_{n,\uparrow \uparrow }\left( t_1\right) \left| n+1,\downarrow \right\rangle
+\widehat{{\cal C}}_{n+1,\downarrow \uparrow }\left( t_3\right) \widehat{%
{\cal J}}_{n\uparrow \downarrow }\left( t_2\right) \widehat{{\cal C}}%
_{n,\uparrow \uparrow }\left( t_1\right) \left| n+1,\uparrow \right\rangle \\
&&+\widehat{{\cal C}}_{n-1,\uparrow \uparrow }\left( t_3\right) \widehat{%
{\cal J}}_{n\downarrow \uparrow }\left( t_2\right) \widehat{{\cal C}}%
_{n,\uparrow \downarrow }\left( t_1\right) \left| n-1,\uparrow \right\rangle
+\widehat{{\cal C}}_{n-1,\uparrow \downarrow }\left( t_3\right) \widehat{%
{\cal J}}_{n\downarrow \uparrow }\left( t_2\right) \widehat{{\cal C}}%
_{n,\uparrow \downarrow }\left( t_1\right) \left| n-1,\downarrow
\right\rangle \\
&&\left. +\left( \widehat{{\cal C}}_{n,\downarrow \uparrow }\left(
t_3\right) \widehat{{\cal J}}_{n,\downarrow \downarrow }\left( t_2\right) 
\widehat{{\cal C}}_{n,\uparrow \downarrow }\left( t_1\right) +\widehat{{\cal %
C}}_{n,\uparrow \uparrow }\left( t_3\right) \widehat{{\cal J}}_{n,\uparrow
\uparrow }\left( t_2\right) \widehat{{\cal C}}_{n,\uparrow \uparrow }\left(
t_1\right) \right) \left| n,\uparrow \right\rangle \right] \otimes \left| 
{\bf C}_2,{\bf J},{\bf C}_1\right\rangle \\
&=&\left| \psi ^{^{III}}\right\rangle .
\end{eqnarray*}
with $t_3$ and $\varphi _3$ standing for the duration and phase of the
second $C$ laser pulse. In the absence of fluorescence signal, the above
entangled state $\left| \psi ^{^{III}}\right\rangle $ evolves to

\begin{eqnarray}
\left| \psi ^{^{IV}}\right\rangle &=&{\sf N}\sum_{n=0}^\infty \left[ \Lambda
_{n-1}^{(0)}\widehat{{\cal C}}_{n,\downarrow \uparrow }\left( t_3\right) 
\widehat{{\cal J}}_{n-1,\uparrow \downarrow }\left( t_2\right) \widehat{%
{\cal C}}_{n-1,\uparrow \uparrow }\left( t_1\right) \left( 1-\delta
_{n,0}\right) +\right.  \nonumber \\
&&+\Lambda _{n+1}^{(0)}\widehat{{\cal C}}_{n,\uparrow \uparrow }\left(
t_3\right) \widehat{{\cal J}}_{n+1\downarrow \uparrow }\left( t_2\right) 
\widehat{{\cal C}}_{n+1,\uparrow \downarrow }\left( t_1\right) +\Lambda
_n^{(0)}\widehat{{\cal C}}_{n,\downarrow \uparrow }\left( t_3\right) 
\widehat{{\cal J}}_{n,\downarrow \downarrow }\left( t_2\right) \widehat{%
{\cal C}}_{n,\uparrow \downarrow }\left( t_1\right) +  \nonumber \\
&&\left. +\Lambda _n^{(0)}\widehat{{\cal C}}_{n,\uparrow \uparrow }\left(
t_3\right) \widehat{{\cal J}}_{n,\uparrow \uparrow }\left( t_2\right) 
\widehat{{\cal C}}_{n,\uparrow \uparrow }\left( t_1\right) \right] \left|
n\uparrow \right\rangle \otimes \left| {\bf C}_2,{\bf J},{\bf C}%
_1\right\rangle ,  \label{psi4}
\end{eqnarray}
\ and tracing out the auxiliary spaces ${\bf C}_2,{\bf J}$ and ${\bf C}_1$,
the ionic reduced density matrix follows:

\begin{equation}
\widehat{\rho }_{ion}(t_1,t_2,t_3)={\rm {Tr}}_{\zeta _{2,}\xi ,\zeta
_1}\left| \psi ^{^{IV}}\right\rangle \left\langle \psi ^{^{IV}}\right| ={\sf %
N}^2\sum_{n,m=0}^\infty \rho _{n,m}(t_1,t_2,t_3)\left| n\uparrow
\right\rangle \left\langle m\uparrow \right| ,
\end{equation}
where the coefficients $\rho _{n,m}$ are given in Appendix B and the
normalization constant reads ${\sf N}=\left( \sum_{n=0}^\infty \rho
_{n,n}(t_1,t_2,t_3)\right) ^{-1/2}$. The probability of detecting absence of
fluorescence is ${\cal P}=1/{\sf N}^2$ \ and the fidelity${\cal \ }$of the
sculpted mixed state is given by

\begin{equation}
{\cal F}=\left\langle \Psi _d\right| \widehat{\rho }_{ion}(t_1,t_2,t_3)%
\left| \Psi _d\right\rangle .  \label{fidelf}
\end{equation}
We stress that in the present situation we do not have a system of nonlinear
equations as in (\ref{7a}). However, the best parameters (average excitation
of initial coherent state, interaction times and phases of the pulses) to
reach the desired state are determined by numerical maximization of the
fidelity expression in Eq. (\ref{fidelf}), as shown below.

\section{Algorithm for Optimizing the Fidelity of a Sculpted State}

We begin by stressing that there is a crucial difference between the
sculpture scheme for the ideal case, described in Sec. III, and for the
realistic situation with the effects of noise, described in the previous
section. In fact, in the protocol proposed in Sec. III, it is necessary to
solve a nonlinear system of equations for $\beta _k$ and $\varepsilon _k$,
which arise from the recurrence equations relating the amplitudes $\Lambda
_k^{(n)}$. On the other hand, when taking into account the intensity
fluctuations in the laser pulses, there is no immediate way to extract
useful information from a recurrence relation for the operators $\widehat{%
{\cal C}}_{n,jk}\left( t_1\right) $, {\bf \ } $\widehat{{\cal J}}%
_{n,jk}\left( t_2\right) $,\ and $\widehat{{\cal C}}_{n,jk}\left( t_3\right) 
$ ($n=0,1,2....$, $j=\uparrow ,\downarrow $\ and $k=\uparrow ,\downarrow $)
associated with each required cycle. Thus, the first question that arises is
how can this problem be overcome? We propose the following solution:
consider a specific desired motional state to be sculpted and calculate the
fidelity of the sculpted mixed state (\ref{fidelf}). The resulting
expression for the fidelity is considerably involved and a numerical
calculation must be performed to obtain the values $t_i$ and $\varphi _i$ ($%
i=1,2,3)$ that maximize the fidelity-probability rate ${\cal R}\equiv {\cal F%
}^\xi .{\cal P}^\zeta $. Starting then from the average excitation $%
\overline{n}_\alpha $ that leads to the highest fidelity deduced in the
ideal case, we proceed by numerical optimization to find the value of $%
\overline{n}_\alpha $ which maximizes the rate ${\cal R}${\bf . }Following
this procedure, to sculpt the phase state (\ref{desejado}) under the effects
of noise, we find that{\bf \ }$\overline{n}_\alpha =0.25$\ (coincidently the
same value as in the ideal case) results in a highest rate ${\cal R}=0.64$,
which follows from a fidelity ${\cal F}=0.91$\ and probability ${\cal P}%
=0.86 $ of measuring absence of fluorescence.

Fig.4(a) displays the Wigner distribution function for the desired state (%
\ref{desejado}), which is to be compared with the Wigner function in
Fig.3(b) for the sculpted state in the ideal case, with fidelity ${\cal F}%
=0.99$. Fig.4(b) displays the mixed state sculpted in the presence of noise
using the same parameters $\beta _1$ and $\varepsilon _1$ of Fig.3(b). By
comparing both figures we see that the interference terms leading to
negative contributions in Fig.4(b) have diffused into each other and
practically cancel out in Fig.4(b), which correspond to ${\cal F}=0.85$ and $%
{\cal P}=0.40$. Finally, Fig.4(c) displays the mixed state sculpted through
our numerical optimization procedure. The fidelity and probability, ${\cal F}%
=0.91$ and ${\cal P}=0.86$, respectively, resulting from our optimization
procedure are considerably higher than those obtained for the mixed state in
Fig.4(b). It is worth noting that although the Wigner distribution in
Fig.4(b) seems to be closely to the distribution in Fig.4(a) than that in
Fig.4(c), the optimized mixed state possesses a higher fidelity, which is
just a measure of the distance between the vectors representing the desired
and obtained states in Hilbert space and can not be inferred from the Wigner
function (see discussion in Appendix C).

To plot Figs.4(a)-(c), we have used realistic values \cite{wine} for the
Lamb-Dicke parameter, $\eta \simeq 0.202$, Rabi frequency, $\Omega /2\pi
\simeq 475$\ kHz, and we assume the parameter $\Gamma \approx 10^{-8}$s,
estimated in \cite{milburn98} to obtain a good agreement with the
experimental results in Ref \cite{wine}.

\section{Conclusion}

In this work we present a technique to engineer arbitrary ionic motional
states employing projection synthesis. This method consists in sculpt the
desired motional state from a coherent motional state previously prepared in
the ionic trap. The sculpture of arbitrary states from a coherent
superposition by projection synthesis was previously developed in the cavity
QED domain \cite{serra}. However, here we take advantage of the facilities
in manipulating trapped ions to make the sculpture process even more
attractive for experimental implementation. As in the cavity QED context,
instead of requiring $N$ laser pulses to generate an arbitrary motional
state with a maximum number of phonons equal to $N$ \cite{kneer}, our
technique utilizes just $M=int\left[ \left( N+1\right) /2\right] $ pulses.

The sculpture scheme was also developed for the realistic situation of
intensity fluctuations in the required laser pulses. In this connection, a
phenomenological-operator approach developed recently to account for errors
in complex quantum processes was applied \cite{poa,norton}. This approach
furnishes a straightforward technique to estimate the fidelity resulting
from the engineering process, withoud the need to perform the usual extended 
{\it ab initio} calculations required by standard methods. The reasoning
behind the phenomenological approach developed here is to incorporate in a
concise algebraic form the main results obtained from the application of the
master equation (to the effects of noise arising from the intensity
fluctuations in the laser pulses) as done in Ref. \cite{milburn98}. By
defining an auxiliary state space where phenomenological operators are
defined, we account for the effects of noise explicitly in the evolution of
the state vector of the whole system comprehended by the ionic and auxiliary
states{\bf . }After computing the evolved state of the whole system, we can
immediately obtain the reduced density matrix of the ionic system by tracing
out the auxiliary variables. We stress that the results obtained here from
the phenomenological procedure are quite general, and in principle can be
applied to any quantum process in an ionic trap, such as quantum
communication, logic operations and teleportation.

As an application of the technique for sculpting an arbitrary motional ionic
state, combined with the phenomenological approach to the analysis of the
effects of noise in the process, we have computed the fidelity for sculpting
a truncated phase state. We assumed realistic values for the parameters
involved. Moreover, we have proposed an algorithm to optimize the fidelity
of a sculpted state in the presence of noise. This algorithm consists in
maximizing the fidelity of the sculpted state (subjected to noise), with
regard to the desired state through a convenient choice of the parameters
involved in the process. We have also shown, through the phenomenological
approach, that the noise introduced by the Jaynes-Cummings pulse has less
effect than that coming from the carrier pulse for a number of phonons $n$ $%
\lesssim 25$. Thus, whether for sculpting a motional state or for any other
process involving a small number of phonons, such as engineering a motional
qubit state $c_0\left| 0\right\rangle +c_1\left| 1\right\rangle $ for
quantum computation, as performed in \cite{CNOT}, the main source of errors
will undoubtedly be the carrier pulses.

\begin{center}
{\bf Acknowledgments}
\end{center}

We wish to thank the CNPq and FAPESP, Brazilian agencies.

{\huge Appendix A}

In this appendix we show explicitly the elements $\left\langle {\bf C}\left| 
\widehat{{\cal C}}_{n,jk}\left( t,\varphi \right) \widehat{{\cal C}}%
_{n^{\shortmid }j^{\shortmid }k^{\shortmid }}^{\dagger }\left( t,\varphi
\right) \right| {\bf C}\right\rangle $\ ($n,n^{\shortmid }=0,1,2....$, $j,$ $%
j^{\shortmid }=\uparrow ,\downarrow $ and $k,k^{\shortmid }=\uparrow
,\downarrow $) for the carrier pulse.

\begin{eqnarray*}
\left\langle \widehat{{\cal C}}_{n,\downarrow \downarrow }\left( t,\varphi
\right) \widehat{{\cal C}}_{m,\downarrow \downarrow }^{\dagger }\left(
t,\varphi \right) \right\rangle _{{\bf C}} &=&\left\langle \widehat{{\cal C}}%
_{n,\uparrow \uparrow }\left( t,\varphi \right) \widehat{{\cal C}}%
_{m,\uparrow \uparrow }^{\dagger }\left( t,\varphi \right) \right\rangle _{%
{\bf C}}=\frac 12\left[ 1+\cos (2\Omega t)e^{-2\Gamma \Omega ^2t}\right] , \\
\left\langle \widehat{{\cal C}}_{n,\uparrow \downarrow }\left( t,\varphi
\right) \widehat{{\cal C}}_{m,\uparrow \downarrow }^{\dagger }\left(
t,\varphi \right) \right\rangle _{{\bf C}} &=&\left\langle \widehat{{\cal C}}%
_{n,\downarrow \uparrow }\left( t,\varphi \right) \widehat{{\cal C}}%
_{m,\downarrow \uparrow }^{\dagger }\left( t,\varphi \right) \right\rangle _{%
{\bf C}}=\frac 12\left[ 1-\cos (2\Omega t)e^{-2\Gamma \Omega ^2t}\right] , \\
\left\langle \widehat{{\cal C}}_{n,\downarrow \downarrow }\left( t,\varphi
\right) \widehat{{\cal C}}_{m,\downarrow \uparrow }^{\dagger }\left(
t,\varphi \right) \right\rangle _{{\bf C}} &=&\left\langle \widehat{{\cal C}}%
_{n,\downarrow \uparrow }\left( t,\varphi \right) \widehat{{\cal C}}%
_{m,\downarrow \downarrow }^{\dagger }\left( t,\varphi \right) \right\rangle
_{{\bf C}}^{*}=i\frac{e^{-i\varphi }}2\sin (2\Omega t)e^{-2\Gamma \Omega
^2t}, \\
\left\langle \widehat{{\cal C}}_{n,\uparrow \uparrow }\left( t,\varphi
\right) \widehat{{\cal C}}_{m,\uparrow \downarrow }^{\dagger }\left(
t,\varphi \right) \right\rangle _{{\bf C}} &=&\left\langle \widehat{{\cal C}}%
_{n,\uparrow \downarrow }\left( t,\varphi \right) \widehat{{\cal C}}%
_{m,\uparrow \uparrow }^{\dagger }\left( t,\varphi \right) \right\rangle _{%
{\bf C}}^{*}=i\frac{e^{i\varphi }}2\sin (2\Omega t)e^{-2\Gamma \Omega ^2t},
\\
\left\langle \widehat{{\cal C}}_{n,\downarrow \downarrow }\left( t,\varphi
\right) \widehat{{\cal C}}_{m,\uparrow \downarrow }^{\dagger }\left(
t,\varphi \right) \right\rangle _{{\bf C}} &=&\left\langle \widehat{{\cal C}}%
_{n,\uparrow \downarrow }\left( t,\varphi \right) \widehat{{\cal C}}%
_{m,\downarrow \downarrow }^{\dagger }\left( t,\varphi \right) \right\rangle
_{{\bf C}}^{*}=i\frac{e^{i\varphi }}2\sin \left( 2\Omega t\right)
e^{-2\Gamma \Omega ^2t}, \\
\left\langle \widehat{{\cal C}}_{n,\uparrow \downarrow }\left( t,\varphi
\right) \widehat{{\cal C}}_{m,\downarrow \uparrow }^{\dagger }\left(
t,\varphi \right) \right\rangle _{{\bf C}} &=&\left\langle \widehat{{\cal C}}%
_{n,\downarrow \uparrow }\left( t,\varphi \right) \widehat{{\cal C}}%
_{m,\uparrow \downarrow }^{\dagger }\left( t,\varphi \right) \right\rangle _{%
{\bf C}}^{*}=\frac{e^{-2i\varphi }}2\left( 1-\cos \left( 2\Omega t\right)
e^{-2\Gamma \Omega ^2t}\right) , \\
\left\langle \widehat{{\cal C}}_{n,\downarrow \downarrow }\left( t,\varphi
\right) \widehat{{\cal C}}_{m,\uparrow \uparrow }^{\dagger }\left( t,\varphi
\right) \right\rangle _{{\bf C}} &=&\left\langle \widehat{{\cal C}}%
_{n,\uparrow \uparrow }\left( t,\varphi \right) \widehat{{\cal C}}%
_{m,\downarrow \downarrow }^{\dagger }\left( t,\varphi \right) \right\rangle
_{{\bf C}}=\frac 12\left( 1+\cos \left( 2\Omega t\right) e^{-2\Gamma \Omega
^2t}\right) , \\
\left\langle \widehat{{\cal C}}_{n,\downarrow \uparrow }\left( t,\varphi
\right) \widehat{{\cal C}}_{m,\uparrow \uparrow }^{\dagger }\left( t,\varphi
\right) \right\rangle _{{\bf C}} &=&\left\langle \widehat{{\cal C}}%
_{n,\uparrow \uparrow }\left( t,\varphi \right) \widehat{{\cal C}}%
_{m,\downarrow \uparrow }^{\dagger }\left( t,\varphi \right) \right\rangle _{%
{\bf C}}^{*}=-i\frac{e^{i\varphi }}2\sin \left( 2\Omega t\right) e^{-2\Gamma
\Omega ^2t}.
\end{eqnarray*}
For the Jaynes Cummings pulse the elements $\left\langle {\bf J}\left| 
\widehat{{\cal J}}_{n,jk}\left( t,\varphi \right) \widehat{{\cal J}}%
_{n^{\shortmid },j^{\shortmid }k^{\shortmid }}^{\dagger }\left( t,\varphi
\right) \right| {\bf J}\right\rangle $ are

\begin{eqnarray*}
\left\langle \widehat{{\cal J}}_{n,\downarrow \downarrow }^{\dagger }\left(
t,\varphi \right) \widehat{{\cal J}}_{m,\downarrow \downarrow }^{\dagger
}\left( t,\varphi \right) \right\rangle _{{\bf J}} &=&\left\langle \widehat{%
{\cal J}}_{n-1,\uparrow \uparrow }\left( t,\varphi \right) \widehat{{\cal J}}%
_{m-1,\uparrow \uparrow }^{\dagger }\left( t,\varphi \right) \right\rangle _{%
{\bf J}}=\frac{1}{2}\left[ A_{n,m}+B_{n,m}\right] , \\
\left\langle \widehat{{\cal J}}_{n,\downarrow \downarrow }\left( t,\varphi
\right) \widehat{{\cal J}}_{m-1,\uparrow \downarrow }^{\dagger }\left(
t,\varphi \right) \right\rangle _{{\bf J}} &=&\left\langle \widehat{{\cal J}}%
_{m,\uparrow \downarrow }\left( t,\varphi \right) \widehat{{\cal J}}%
_{n-1,\downarrow \downarrow }^{\dagger }\left( t,\varphi \right)
\right\rangle _{{\bf J}}^{\ast }=\frac{e^{i\varphi }}{2}\left[
C_{n,m}-D_{n,m}\right] , \\
\left\langle \widehat{{\cal J}}_{n-1,\uparrow \downarrow }\left( t,\varphi
\right) \widehat{{\cal J}}_{m-1,\uparrow \downarrow }^{\dagger }\left(
t,\varphi \right) \right\rangle _{{\bf J}} &=&\left\langle \widehat{{\cal J}}%
_{n,\downarrow \uparrow }\left( t,\varphi \right) \widehat{{\cal J}}%
_{m,\downarrow \uparrow }^{\dagger }\left( t,\varphi \right) \right\rangle _{%
{\bf J}}=\frac{1}{2}\left[ A_{n,m}-B_{n,m}\right] , \\
\left\langle \widehat{{\cal J}}_{n,\downarrow \downarrow }\left( t,\varphi
\right) \widehat{{\cal J}}_{m+1,\downarrow \uparrow }^{\dagger }\left(
t,\varphi \right) \right\rangle _{{\bf J}} &=&\left\langle \widehat{{\cal J}}%
_{m,\downarrow \uparrow }\left( t,\varphi \right) \widehat{{\cal J}}%
_{n+1,\downarrow \downarrow }^{\dagger }\left( t,\varphi \right)
\right\rangle _{{\bf J}}^{\ast }=\frac{e^{-i\varphi }}{2}\left[
-C_{n,m+1}+D_{n,m+1}\right] , \\
\left\langle \widehat{{\cal J}}_{n,\downarrow \downarrow }\left( t,\varphi
\right) \widehat{{\cal J}}_{m,\uparrow \uparrow }^{\dagger }\left( t,\varphi
\right) \right\rangle _{{\bf J}} &=&\left\langle \widehat{{\cal J}}%
_{m,\uparrow \uparrow }\left( t,\varphi \right) \widehat{{\cal J}}%
_{n,\downarrow \downarrow }^{\dagger }\left( t,\varphi \right) \right\rangle
_{{\bf J}}=\frac{1}{2}\left[ A_{n,m+1}+B_{n,m+1}\right] , \\
\left\langle \widehat{{\cal J}}_{n-1,\uparrow \downarrow }\left( t,\varphi
\right) \widehat{{\cal J}}_{m+1,\downarrow \uparrow }^{\dagger }\left(
t,\varphi \right) \right\rangle _{{\bf J}} &=&\left\langle \widehat{{\cal J}}%
_{m-1,\downarrow \uparrow }\left( t,\varphi \right) \widehat{{\cal J}}%
_{n+1,\uparrow \downarrow }^{\dagger }\left( t,\varphi \right) \right\rangle
_{{\bf J}}^{\ast }=\frac{e^{-2i\varphi }}{2}\left[ -A_{n,m+1}+B_{n,m+1}%
\right] , \\
\left\langle \widehat{{\cal J}}_{n-1,\uparrow \downarrow }\left( t,\varphi
\right) \widehat{{\cal J}}_{m,\uparrow \uparrow }^{\dagger }\left( t,\varphi
\right) \right\rangle _{{\bf J}} &=&\left\langle \widehat{{\cal J}}%
_{m-1,\uparrow \uparrow }\left( t,\varphi \right) \widehat{{\cal J}}%
_{n,\uparrow \downarrow }^{\dagger }\left( t,\varphi \right) \right\rangle _{%
{\bf J}}^{\ast }=-\frac{e^{-i\varphi }}{2}\left[ C_{n,m+1}+D_{n,m+1}\right] ,
\\
\left\langle \widehat{{\cal J}}_{n+1,\downarrow \uparrow }\left( t,\varphi
\right) \widehat{{\cal J}}_{m,\uparrow \uparrow }^{\dagger }\left( t,\varphi
\right) \right\rangle _{{\bf J}} &=&\left\langle \widehat{{\cal J}}%
_{m+1,\uparrow \uparrow }\left( t,\varphi \right) \widehat{{\cal J}}%
_{n,\downarrow \uparrow }^{\dagger }\left( t,\varphi \right) \right\rangle _{%
{\bf J}}^{\ast }=-\frac{e^{i\varphi }}{2}\left[ C_{n+1,m+1}+D_{n+1,m+1}%
\right] ,
\end{eqnarray*}
where

\begin{eqnarray*}
A_{n,m} &=&\cos \left[ gt\left( \sqrt{n}-\sqrt{m}\right) \right] \exp \left[
-\Gamma g^{2}t\left( \sqrt{n}-\sqrt{m}\right) ^{2}/2\right] {\rm {,}} \\
B_{n,m} &=&\cos \left[ gt\left( \sqrt{n}+\sqrt{m}\right) \right] \exp \left[
-\Gamma g^{2}t\left( \sqrt{n}+\sqrt{m}\right) ^{2}/2\right] {\rm {,}} \\
C_{n,m} &=&\sin \left[ gt\left( \sqrt{n}-\sqrt{m}\right) \right] \exp \left[
-\Gamma g^{2}t\left( \sqrt{n}-\sqrt{m}\right) ^{2}/2\right] {\rm {,}}
\end{eqnarray*}
and 
\[
D_{n,m}=\sin \left[ gt\left( \sqrt{n}+\sqrt{m}\right) \right] \exp \left[
-\Gamma g^{2}t\left( \sqrt{n}+\sqrt{m}\right) ^{2}/2\right] {\rm {.}} 
\]

{\huge Appendix B}

In this appendix we show explicitly the elements $\rho
_{n,m}(t_{1},t_{2},t_{3})$

\begin{eqnarray*}
\rho _{n,m}(t_1,t_2,t_3) &=&\Lambda _{n-1}^{(0)}\left( \Lambda
_{m-1}^{(0)}\right) ^{*}\frac 18\left[ 1-\cos \left( 2\Omega t_3\right)
e^{-2\Gamma \Omega ^2t_3}\right] \times \\
&&\times \left[ A_{n,m}\left( t_2\right) -B_{n,m}\left( t_2\right) \right] %
\left[ 1+\cos \left( 2\Omega t_1\right) e^{-2\Gamma \Omega ^2t_1}\right]
\left( 1-\delta _{n,0}\right) \left( 1-\delta _{m,0}\right) + \\
&&-\Lambda _{n-1}^{(0)}\left( \Lambda _{m+1}^{(0)}\right) ^{*}\frac{%
e^{i\varphi _3}}8\sin \left( 2\Omega t_3\right) e^{-2\Gamma \Omega
^2t_3}\times \\
&&\times e^{-2i\varphi _2}\left[ -A_{n,m+1}\left( t_2\right)
+B_{n,m+1}\left( t_2\right) \right] e^{i\varphi _1}\sin \left( 2\Omega
t_1\right) e^{-2\Gamma \Omega ^2t_1}\left( 1-\delta _{n,0}\right) + \\
&&+\Lambda _{n-1}^{(0)}\left( \Lambda _m^{(0)}\right) ^{*}\frac i8\left[
1-\cos \left( 2\Omega t_3\right) e^{-2\Gamma \Omega ^2t_3}\right] \times \\
&&\times e^{-i\varphi _2}\left[ -C_{n,m}\left( t_2\right) -D_{n,m}\left(
t_2\right) \right] e^{i\varphi _1}\sin \left( 2\Omega t_1\right) e^{-2\Gamma
\Omega ^2t_1}\left( 1-\delta _{n,0}\right) + \\
&&+\Lambda _{n-1}^{(0)}\left( \Lambda _m^{(0)}\right) ^{*}i\frac{e^{i\varphi
_3}}8\sin \left( 2\Omega t_3\right) e^{-2\Gamma \Omega ^2t_3}\times \\
&&\times e^{-i\varphi _2}\left[ C_{n,m+1}\left( t_2\right) +D_{n,m+1}\left(
t_2\right) \right] \left[ 1+\cos \left( 2\Omega t_1\right) e^{-2\Gamma
\Omega ^2t_1}\right] \left( 1-\delta _{n,0}\right) + \\
&&+\Lambda _{n+1}^{(0)}\left( \Lambda _{m-1}^{(0)}\right) ^{*}\frac{%
e^{-i\varphi _3}}8\sin \left( 2\Omega t_3\right) e^{-2\Gamma \Omega
^2t_3}\times \\
&&\times e^{+2i\varphi _2}\left[ -A_{n+1,m}\left( t_2\right)
+B_{n+1,m}\left( t_2\right) \right] e^{-i\varphi _1}\sin \left( 2\Omega
t_1\right) e^{-2\Gamma \Omega ^2t_1}\left( 1-\delta _{m,0}\right) + \\
&&+\Lambda _{n+1}^{(0)}\left( \Lambda _{m+1}^{(0)}\right) ^{*}\frac 18\left[
1+\cos \left( 2\Omega t_3\right) e^{-2\Gamma \Omega ^2t_3}\right] \times \\
&&\times \left[ A_{n+1,m+1}\left( t_2\right) -B_{n+1,m+1}\left( t_2\right) %
\right] \left[ 1-\cos \left( 2\Omega t_1\right) e^{-2\Gamma \Omega ^2t_1}%
\right] + \\
&&+\Lambda _{n+1}^{(0)}\left( \Lambda _m^{(0)}\right) ^{*}i\frac{%
e^{-i\varphi _3}}8\sin \left( 2\Omega t_3\right) e^{-2\Gamma \Omega
^2t_3}\times \\
&&\times e^{i\varphi _2}\left[ C_{n+1,m}\left( t_2\right) +D_{n+1,m}\left(
t_2\right) \right] \left[ 1-\cos \left( 2\Omega t_1\right) e^{-2\Gamma
\Omega ^2t_1}\right] + \\
&&-\Lambda _{n+1}^{(0)}\left( \Lambda _m^{(0)}\right) ^{*}\frac i8\left[
1+\cos \left( 2\Omega t_3\right) e^{-2\Gamma \Omega ^2t_3}\right] \times \\
&&\times e^{i\varphi _2}\left[ C_{n+1,m+1}\left( t_2\right)
+D_{n+1,m+1}\left( t_2\right) \right] e^{-i\varphi _1}\sin \left( 2\Omega
t_1\right) e^{-2\Gamma \Omega ^2t_1} \\
&&-\Lambda _n^{(0)}\left( \Lambda _{m-1}^{(0)}\right) ^{*}\frac i8\left[
1-\cos \left( 2\Omega t_3\right) e^{-2\Gamma \Omega ^2t_3}\right] \times \\
&&\times e^{i\varphi _2}\left[ C_{n,m}\left( t_2\right) -D_{n,m}\left(
t_2\right) \right] e^{-i\varphi _1}\sin \left( 2\Omega t_1\right)
e^{-2\Gamma \Omega ^2t_1}\left( 1-\delta _{m,0}\right) + \\
&&-\Lambda _n^{(0)}\left( \Lambda _{m+1}^{(0)}\right) ^{*}i\frac{e^{i\varphi
_3}}8\sin \left( 2\Omega t_3\right) e^{-2\Gamma \Omega ^2t_3}\times \\
&&\times e^{-i\varphi _2}\left[ -C_{n,m+1}\left( t_2\right) +D_{n,m+1}\left(
t_2\right) \right] \left[ 1-\cos \left( 2\Omega t_1\right) e^{-2\Gamma
\Omega ^2t_1}\right] + \\
&&+\Lambda _n^{(0)}\left( \Lambda _m^{(0)}\right) ^{*}\frac 18\left[ 1-\cos
\left( 2\Omega t_3\right) e^{-2\Gamma \Omega ^2t_3}\right] \times \\
&&\times \left[ A_{n,m}\left( t_2\right) +B_{n,m}\left( t_2\right) \right] %
\left[ 1-\cos \left( 2\Omega t_1\right) e^{-2\Gamma \Omega ^2t_1}\right] + \\
&&-\Lambda _n^{(0)}\left( \Lambda _m^{(0)}\right) ^{*}\frac{e^{i\varphi _3}}8%
\sin \left( 2\Omega t_3\right) e^{-2\Gamma \Omega ^2t_3}\times \\
&&\times \left[ A_{n,m+1}\left( t_2\right) +B_{n,m+1}\left( t_2\right) %
\right] e^{-i\varphi _1}\sin \left( 2\Omega t_1\right) e^{-2\Gamma \Omega
^2t_1}+ \\
&&+\Lambda _n^{(0)}\left( \Lambda _{m-1}^{(0)}\right) ^{*}i\frac{%
e^{-i\varphi _3}}8\sin \left( 2\Omega t_3\right) e^{-2\Gamma \Omega
^2t_3}\times \\
&&\times e^{i\varphi _2}\left[ C_{n+1,m}\left( t_2\right) -D_{n+1,m}\left(
t_2\right) \right] \left[ 1+\cos \left( 2\Omega t_1\right) e^{-2\Gamma
\Omega ^2t_1}\right] \left( 1-\delta _{m,0}\right) + \\
&&+\Lambda _n^{(0)}\left( \Lambda _{m+1}^{(0)}\right) ^{*}\frac i8\left[
1+\cos \left( 2\Omega t_3\right) e^{-2\Gamma \Omega ^2t_3}\right] \times \\
&&\times e^{-i\varphi _2}\left[ -C_{n+1,m+1}\left( t_2\right)
+D_{n+1,m+1}\left( t_2\right) \right] e^{i\varphi _1}\sin \left( 2\Omega
t_1\right) e^{-2\Gamma \Omega ^2t_1}+ \\
&&-\Lambda _n^{(0)}\left( \Lambda _m^{(0)}\right) ^{*}\frac{e^{-i\varphi _3}}%
8\sin \left( 2\Omega t_3\right) e^{-2\Gamma \Omega ^2t_3}\times \\
&&\times \left[ A_{n+1,m}\left( t_2\right) +B_{n+1,m}\left( t_2\right) %
\right] e^{i\varphi _1}\sin \left( 2\Omega t_1\right) e^{-2\Gamma \Omega
^2t_1}+
\end{eqnarray*}
\begin{eqnarray*}
&&\qquad \qquad \quad \quad +\Lambda _n^{(0)}\left( \Lambda _m^{(0)}\right)
^{*}\frac 18\left[ 1+\cos \left( 2\Omega t_3\right) e^{-2\Gamma \Omega ^2t_3}%
\right] \times \\
&&\qquad \qquad \quad \quad \times \left[ A_{n+1,m+1}\left( t_2\right)
+B_{n+1,m+1}\left( t_2\right) \right] \left[ 1+\cos \left( 2\Omega
t_1\right) e^{-2\Gamma \Omega ^2t_1}\right] ,
\end{eqnarray*}
where $\varphi _1$, $\varphi _2$, and $\varphi _3$ are the laser pulse
phases for the first carrier, the Jaynes-Cummings and the second carrier
pulses, respectively. Note that, for $\Gamma =0$ we obtain from the above
expressions the expected results without noise.

{\huge Appendix C}

In this appendix we show that a higher fidelity of the sculpted state, with
regard to the desired state, does not implicate that the shape of the Wigner
distribution function of such sculpted state must be closely to that of the
desired state. We start by showing that for a desired state $\left| \Xi
\right\rangle $ there are an infinite set of sculpted state $\left| \Lambda
_\lambda \right\rangle $ with the same fidelity ${\cal F}=\left|
\left\langle \Lambda _\lambda |\Xi \right\rangle \right| ^2$, the label $%
\lambda $ standing for a continuous real parameter or a set of continuous
real parameters to be defined below. In order to simplify our discussion we
consider a desired vibrational state in the two-dimensional Fock space $%
\left\{ \left| 0\right\rangle ,\left| 1\right\rangle \right\} $: $\left| \Xi
\right\rangle =\left( \left| 0\right\rangle +\left| 1\right\rangle \right) /%
\sqrt{2}$. In this case, $\lambda $ reduces to a single continuous parameter
since the set of states displaying the same fidelity ${\cal F}$ (represented
by unit vectors in the so-called Bloch sphere of ${\Bbb R}^3$, in analogy
with spin-$1/2$ states \cite{feynman}), satisfy $\left| \Lambda _\lambda
\right\rangle $ $=%
%TCIMACRO{\limfunc{e}}%
%BeginExpansion
\mathop{\rm e}%
%EndExpansion
\nolimits^{i\phi _0}\left( \lambda \left| 0\right\rangle +%
%TCIMACRO{\limfunc{e}}%
%BeginExpansion
\mathop{\rm e}%
%EndExpansion
\nolimits^{i(\phi _1-\phi _0)}\sqrt{1-\lambda ^2}\left| 1\right\rangle
\right) $, with 
\begin{equation}
\phi =\phi _1-\phi _0=\arccos \frac{{\cal F}-1/2}{\lambda \sqrt{1-\lambda ^2}%
}.  \label{C1}
\end{equation}
The components $r_i=$Tr $\sigma _i\left| \Lambda _\lambda \right\rangle
\left\langle \Lambda _\lambda \right| $ of the states $\left| \Lambda
_\lambda \right\rangle $ in the Bloch sphere, for $i=x,y,z$ and $\sigma _i$
referring to the Pauli pseudo-spin operators, read 
\begin{eqnarray}
r_x &=&2{\cal F}-1,  \nonumber \\
r_y &=&\pm 2\left[ \lambda ^2\left( \lambda ^2-1\right) +({\cal F}-1/2)^2%
\right] ,  \label{C2} \\
r_z &=&2\lambda ^2-1,  \nonumber
\end{eqnarray}
where the signal $+$($-$) in $r_y$ corresponds to positive (negative) values
of $\phi $. Next, choosing ${\cal F}=(2+\sqrt{3})/4\approx 0.933$, it
follows only two states with $\phi =0$ ($r_y=0$): $\left| \Lambda
_{1/2}\right\rangle $ $=\left( \left| 0\right\rangle +\sqrt{3}\left|
1\right\rangle \right) /2$ and $\left| \Lambda _{\sqrt{3}/2}\right\rangle $ $%
=\left( \sqrt{3}\left| 0\right\rangle +\left| 1\right\rangle \right) /2$,
and an infinite set of states with $\phi \neq 0$. All these states, with
Bloch vectors lying in the cone displayed in Fig. 5, present the same
fidelity $\approx 0.933$, but different Wigner distribution functions. Figs.
6(a,b,c) display the Wigner functions for the states $\left| \Xi
\right\rangle $, $\left| \Lambda _{1/2}\right\rangle $ and $\left| \Lambda _{%
\sqrt{3}/2}\right\rangle $, respectively, showing that although $\left|
\Lambda _{1/2}\right\rangle $ and $\left| \Lambda _{\sqrt{3}/2}\right\rangle 
$ present the same fidelity with respect to $\left| \Xi \right\rangle $,
Figs. 6(b,c) exhibit completely different shapes compared to Fig. 6(a). It
is worth mention that the Wigner function varies continuously when $\left|
\Lambda _\lambda \right\rangle $ evolves continuously from $\left| \Lambda
_{1/2}\right\rangle $ to $\left| \Lambda _{\sqrt{3}/2}\right\rangle $
(going, in positive $z$ direction, through the bottom to the top of the cone
in Fig. 5, in clockwise or anticlockwise direction depending on the signal
of $r_y$).

The same analysis holds for the statistical mixtures 
\begin{equation}
\rho _\lambda (\varkappa )=\lambda ^2\left| 0\right\rangle \left\langle
0\right| +\left( 1-\lambda ^2\right) \left| 1\right\rangle \left\langle
1\right| +\varkappa \lambda \sqrt{1-\lambda ^2}\left( 
%TCIMACRO{\limfunc{e}}%
%BeginExpansion
\mathop{\rm e}%
%EndExpansion
\nolimits^{-i\phi }\left| 0\right\rangle \left\langle 1\right| +%
%TCIMACRO{\limfunc{e}}%
%BeginExpansion
\mathop{\rm e}%
%EndExpansion
\nolimits^{i\phi }\left| 1\right\rangle \left\langle 0\right| \right) ,
\label{C3}
\end{equation}
where the parameter $\varkappa $ $\in [0,1]$ accounts for the purity of the
density operator $\rho _\lambda (\varkappa )$, in a way that $\varkappa \neq
1$ imposes that Tr $\left\{ \left[ \rho _\lambda (\varkappa )\right]
^2\right\} <1$. For a given value of $\varkappa $, it follows an infinite
set of statistical mixtures presenting the same fidelity ${\cal F}%
=\left\langle \Xi |\rho _\lambda (\varkappa )|\Xi \right\rangle $ when 
\[
\phi =\arccos \frac{{\cal F}-1/2}{\varkappa \lambda \sqrt{1-\lambda ^2}}. 
\]
From a particular choice of $\varkappa =0.9$, ${\cal F}=0.8$, and $\lambda
=0.7$, we obtain a statistical mixture displaying a Wigner distribution
function, depicted in Fig. 7, whose shape seems to be closely to that in
Fig. 6(a) than those in Figs. 6(b,c), although its fidelity is smaller than $%
0.933$.

When considering the truncated Fock space with more than two dimension, say $%
\left\{ \left| 0\right\rangle ,\left| 1\right\rangle ,\left| 2\right\rangle
\right\} $, it follows a set of two real parameters for describing the set
of states in a sphere of ${\Bbb R}^4$, and so on. Finally, we mention that
our optimization protocol described in Sec.V is based on the maximization of
the above-defined fidelity, the overlap between the sculpted and the desired
state, which represents a particular measure of distance between vectors in
Hilbert space, and can not be inferred from the Wigner distribution function.

%%%%%%%%%%%%%%%%%%%%%%%%%%%%%%%%%%%%%%%%%%%%%%%%%%%%%%%%%%%%%%%%%%%%%%%
%

{\bf Figure Captions}

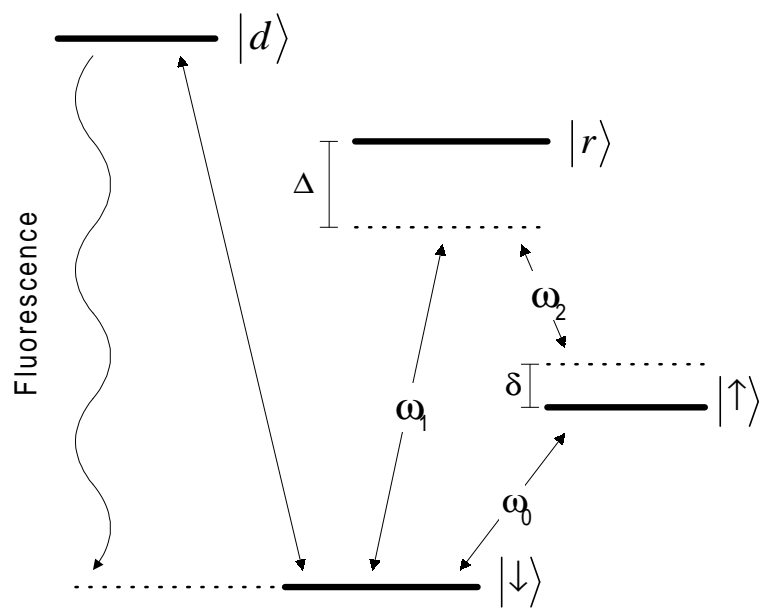
FIG. 1. Electronic energy level diagram of a trapped ion interacting with
the laser beans of frequency $\omega _1$ and $\omega _2$, where $\delta
=\omega _1-\omega _2-\omega _0$ ($\delta \ll \Delta $), $\left|
r\right\rangle $ (adiabatically eliminated) is an auxiliary electronic level
which indirectly couples the levels $\left| \uparrow \right\rangle $ and $%
\left| \downarrow \right\rangle $, and $\left| d\right\rangle $ is an
electronic level used to measure the fluorescence emission .

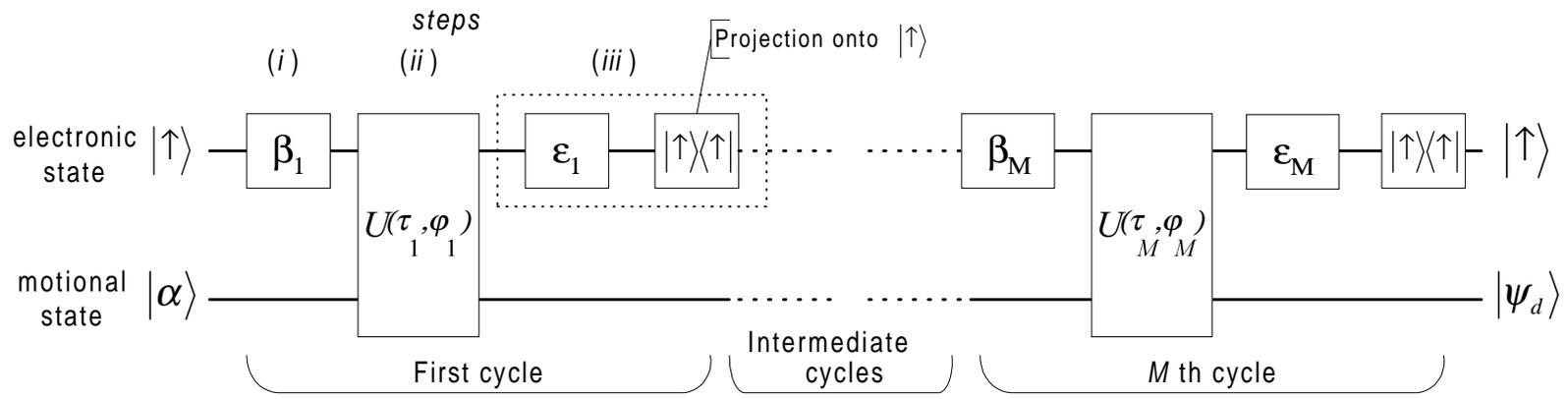
FIG. 2. A quantum algorithm notation \cite{qalg} for the process of
sculpting arbitrary motional states. The complex parameters $\beta _k$ and $%
\varepsilon _k$ indicate rotations of the electronic states by the first and
second $C$ pulses, respectively, in the $k$th cycle and are adjusted by an
appropriate choice of the duration and phase of the laser field following
Eqs. (\ref{rota},\ref{rotb}). $U(\tau _k,\varphi _k)$ indicates the
evolution operator for the $k$th $JC$ pulse. At the end of each cycle the
measurement of absence of fluorescence (projection on to $\left| \uparrow
\right\rangle $) is required for the successful accomplishment of the
engineering process. The time proceeds from left to right, as usual.

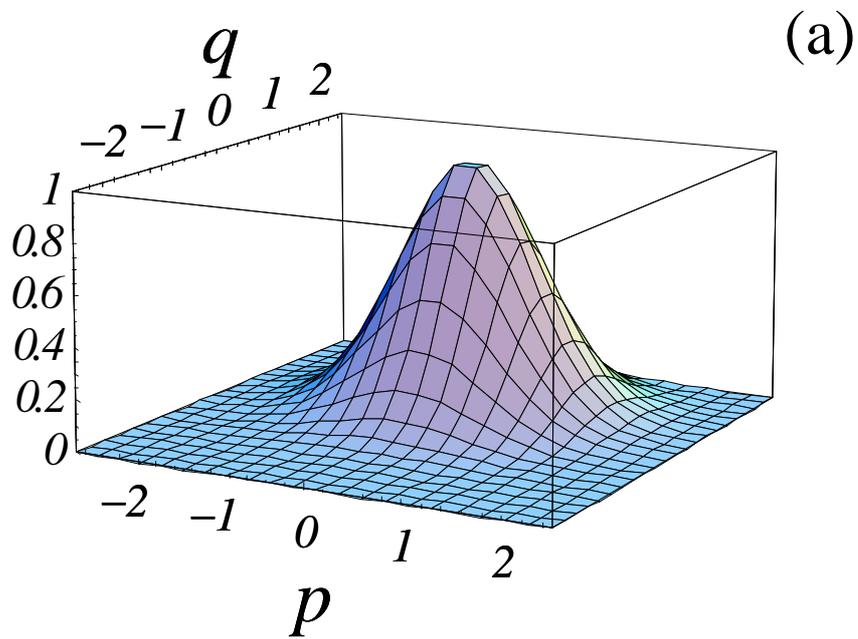
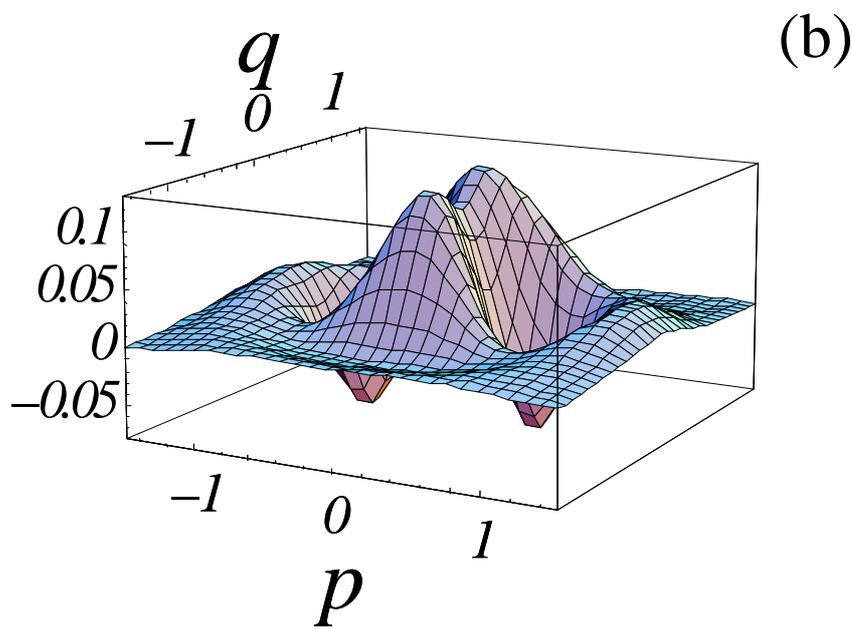
FIG. 3. Wigner distribution functions for (a) the initial coherent state
associated with $\overline{n}_\alpha =0.25$ and for (b) the sculpted state
after the first cycle. We have used the values $\varepsilon
_1=25.6159-I\times 0.0379$ and $\beta _1=-0.3994-I\times 0.6408\times
10^{-4} $ associated with the values in Table 1: ${\cal F}=0.99$, ${\cal P}%
=0.38$ and ${\cal R}=0.60$.

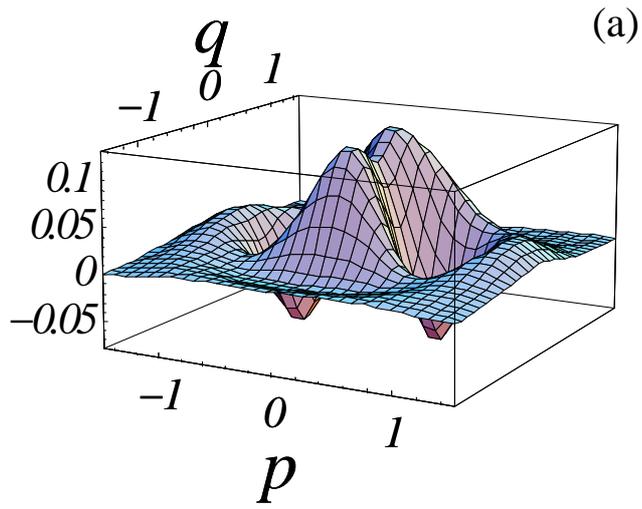
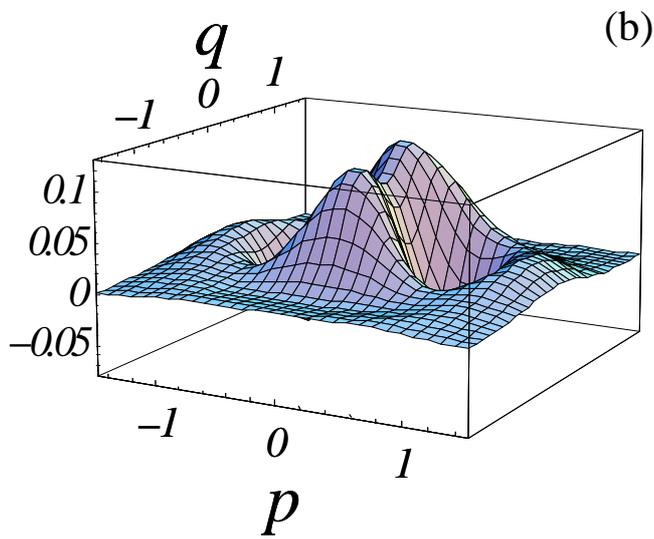
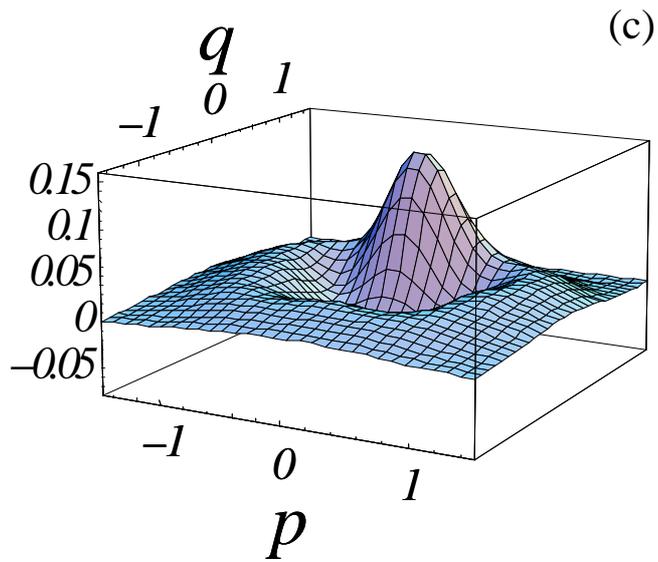
FIG. 4.Wigner distribution functions for (a) the desired phase state given
in Eq. (\ref{desejado}), (b) the mixed state sculpted in the presence of
noise using the same parameters $\beta _1$ and $\varepsilon _1$ of Fig.3(b) (%
${\cal F}=0.85$ and ${\cal P}=0.40$), and (c) the mixed state sculpted
through our numerical optimization procedure using the set of parameters{\bf %
\ }$\Omega t_1=0.56$, $\varphi _1=5.48$, $gt_2=0.75$, $\varphi _2=1.40$, $%
\Omega t_3$ $=1.88$, and $\varphi _3=1.43$ associated with ${\cal F}=0.91,$ $%
{\cal P}=0.86$, and ${\cal R}=0.64$.

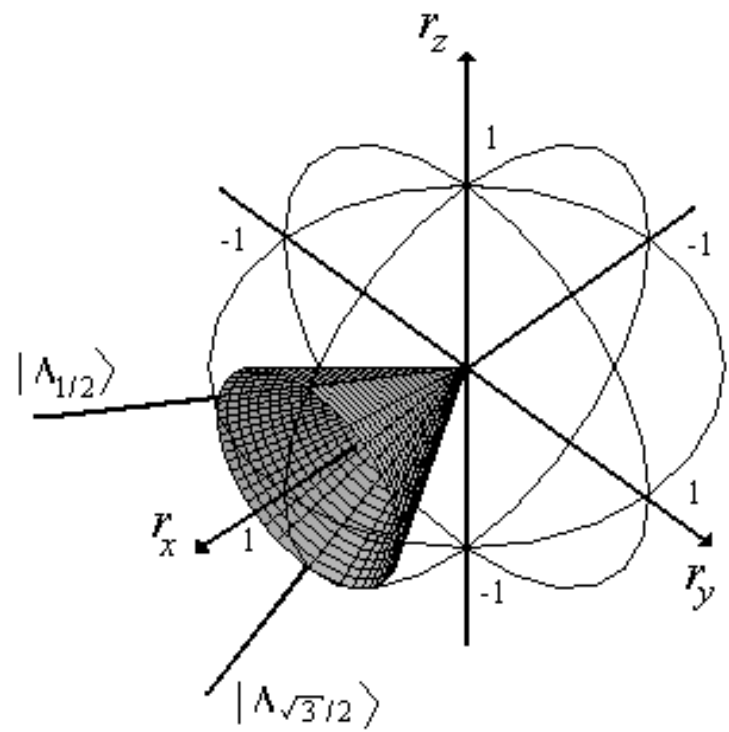
FIG. 5. The cone around the axis $r_x$ displayed in this figure, represents
an infinite set of unit vector states $\left| \Lambda _\lambda \right\rangle 
$ having the same fidelity ${\cal F=}(2+\sqrt{3})/4$ with respect to the
state $\left| \Xi \right\rangle $. The circles around the figure represent
the contours of the unit Bloch sphere.

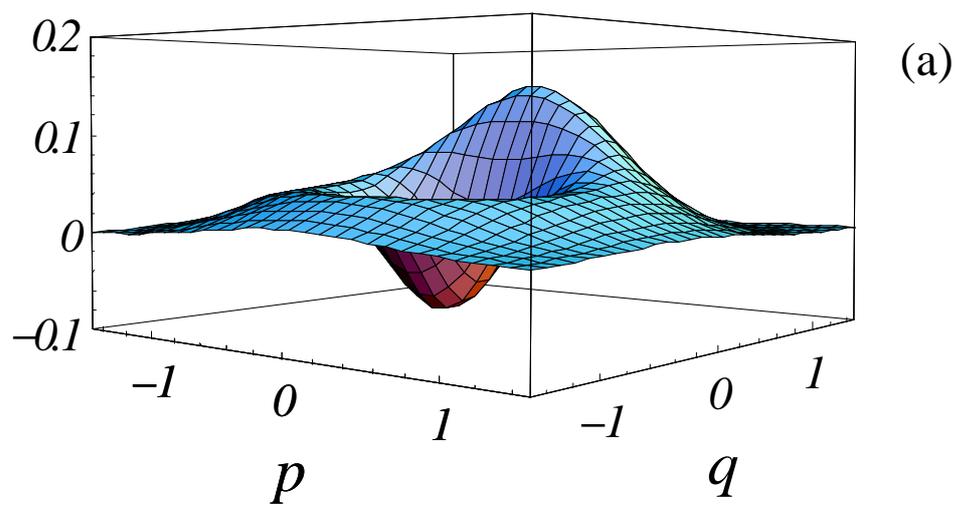
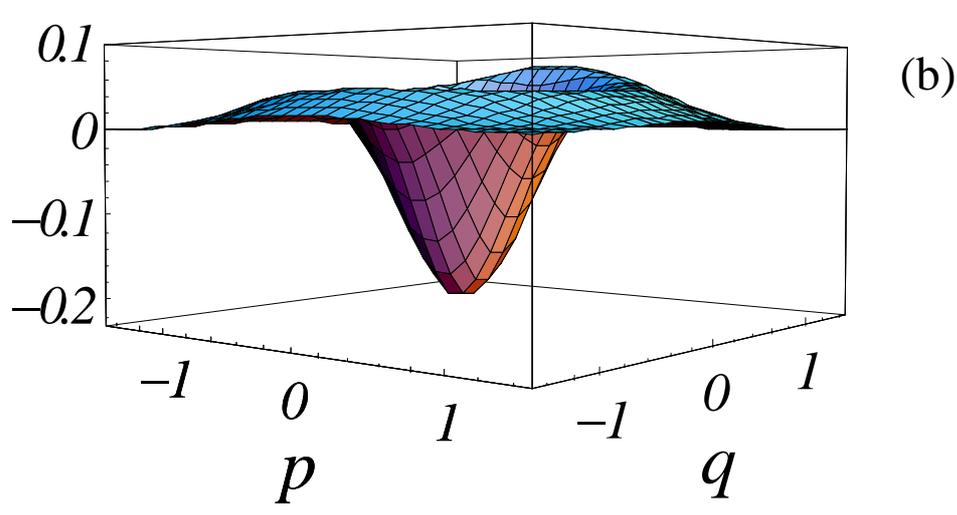
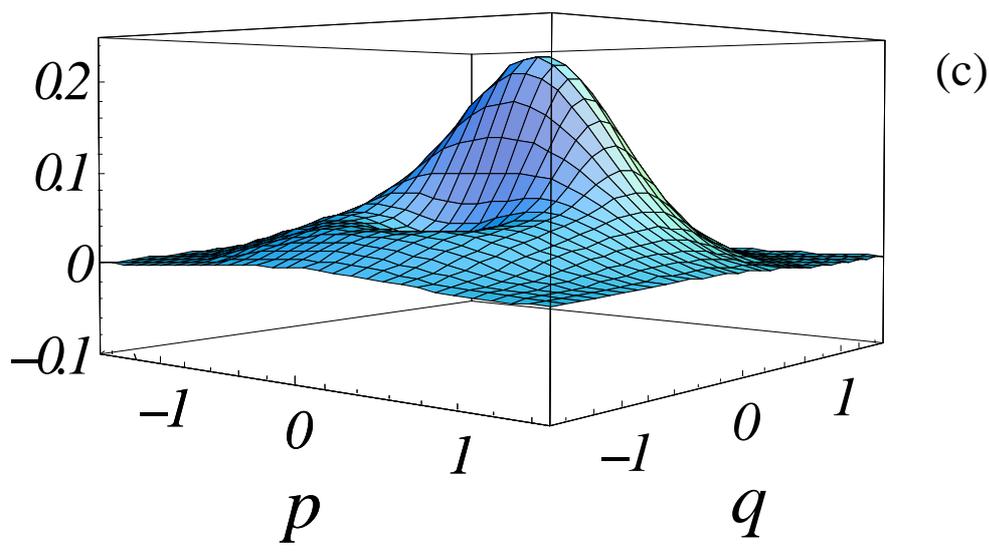
FIG. 6. Wigner distribution functions for (a) the desired state $\left| \Xi
\right\rangle $ and the two states lying in the cone of Fig. 5, with $\phi
=0 $: (b) $\left| \Lambda _{1/2}\right\rangle $ and (c) $\left| \Lambda _{%
\sqrt{3}/2}\right\rangle $.

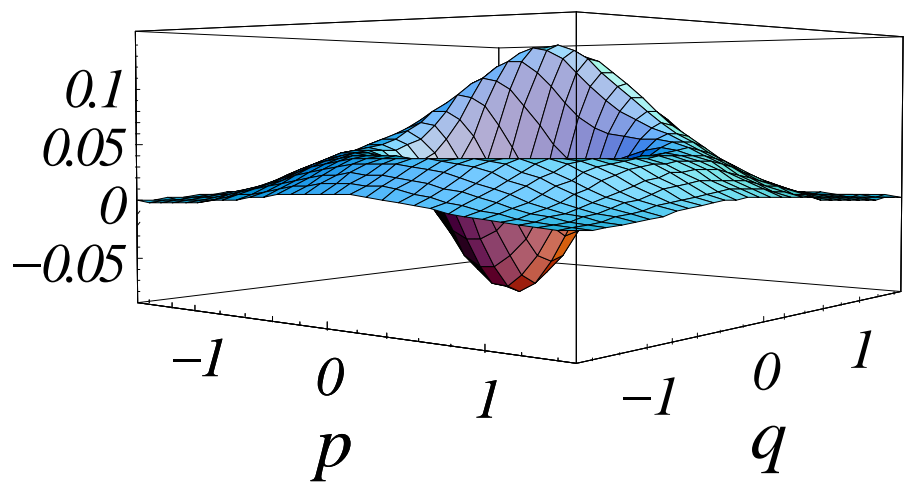
FIG. 7. Wigner distribution function for the statistical mixture $\rho
_\lambda (\varkappa )$ with a particular choice of $\varkappa =0.90$, ${\cal %
F}=0.85$, and $\lambda =0.70$.

{\bf Table}

TABLE 1.The probability ${\cal P}$, fidelity ${\cal F}$, and rate ${\cal R}=%
{\cal F}^\xi .{\cal P}^\zeta $ $\ $(with $\xi =4$ and{\bf \ }$\zeta =1/2$),
for each value of averaged excitation number $\overline{n}_\alpha $. The $JC$
pulse interaction time $g\tau _1$ and phase $\varphi _1$ in this table
correspond to values which maximize the rate ${\cal R}$ for each $\overline{n%
}_\alpha $

\begin{tabular}{llllll}
\hline\hline
$~~\overline{n}_{\alpha }\quad ~$ & $~g\tau _{1}~~$ & $~~\varphi _{1}\quad ~$
& $~{\cal P}~~$ & $~~{\cal F}\quad ~$ & $~{\cal R}~~$ \\ \hline
\end{tabular}

\begin{tabular}{llllll}
$~0.04~$ & $~3.35~$ & $~3.15~$ & $~0.11~$ & $~0.99~$ & $~0.33~$ \\ 
$~0.09$ & $~3.51$ & $~3.14$ & $~0.22$ & $~0.99$ & $~0.47$ \\ 
$~0.16$ & $~3.65$ & $~3.15$ & $~0.33$ & $~0.99$ & $~0.56$ \\ 
$~0.25$ & $~3.79$ & $~3.14$ & $~0.38$ & $~0.99$ & $~0.60$ \\ 
$~0.36$ & $~3.93$ & $~3.14$ & $~0.42$ & $~0.97$ & $~0.59$ \\ 
$~0.49$ & $~4.07$ & $~0.02$ & $~0.44$ & $~0.95$ & $~0.53$ \\ 
$~0.64$ & $~1.81$ & $~3.14$ & $~0.61$ & $~0.92$ & $~0.54$ \\ \hline\hline
\end{tabular}

\end{document}